# Disclosing the Impact of Local Host Effects on TADF Dynamics

Björn Ewald,*[a] Theodor Kaiser,[a] Thomas Fleischmann[a] and Jens Pflaum*[a,b]

Donor-acceptor (D-A) type thermally activated delayed fluorescence (TADF), a key technology of proposed Gen3 organic light emitting diodes (OLEDs), is highly sensitive to the rigidity and polarity of the local environment. Specifically, the torsional flexibility of the D-A dihedral angle and the dipole character of charge transfer states give rise to a distribution of TADF dynamics across the emitter ensemble. Here, we employ single molecule spectroscopy to access the photophysics of individual emitters, thus overcoming the limitations of ensemble averaging. Using photon correlation measurements and locally resolved spectral data from single D-A type TADF molecules embedded in host materials of different polarity and rigidity, we derive host-dependent characteristics and distributions in the TADF dynamics. These are directly linked to local conformational freedom and dielectric properties, offering new insight into host-emitter interactions and enabling rational design strategies for optimized host-emitter combinations in OLED applications.

## Introduction

Thermally activated delayed fluorescence (TADF) is of central interest for next generation organic light emitting diodes (OLEDs), as the mechanism enables efficient triplet exciton harvesting through reverse intersystem crossing (RISC). Generally, TADF molecules consist of a donor (D) and acceptor (A) unit linked via a bridging element, alleviating the spatial overlap of the highest occupied molecular orbital (HOMO) and the lowest unoccupied molecular orbital (LUMO), thus minimizing electronic exchange interaction.[1-3] The $S_1$/$T_1$ states of a D-A emitter constitute charge transfer (CT) states of different spin multiplicity ($^{1}$CT and $^{3}$CT) with inherent excited state dipole moment. A third localized excited (LE) triplet state ($^{3}$LE) typically with $\pi$-$\pi^*$-character and localized on either the D or A moiety, mediates the spin-forbidden ISC/RISC processes via a spin-vibronic coupling mechanism. The RISC rate is dependent on the second-order interaction of the $^{1}$CT and $^{3}$CT states via the $^{3}$LE intermediate state:[4-6]

$$k_{RISC} = \frac{2\pi}{\hbar} \left| \frac{\langle \psi_{1_{CT}} | \widehat{\mathcal{H}}_{SO} | \psi_{3_{LE}} \rangle \langle \psi_{3_{LE}} | \widehat{\mathcal{T}}_{N} | \psi_{3_{CT}} \rangle}{E_{3_{CT}} - E_{3_{LE}}} \right|^2 \delta\left(E_{1_{CT}} - E_{3_{LE}}\right) \qquad (1)$$

Here, $\psi$ are the respective molecular wavefunctions, $E$ the corresponding energies, $\widehat{\mathcal{H}}_{SO}$ is the spin-orbit coupling operator and $\widehat{\mathcal{T}}_{N}$ denotes the vibronic coupling operator. Importantly, the spin-vibronic mechanism is highly sensitive to the local environment, particularly the dielectric and steric properties of the host material.[7-11]

[a] Experimental Physics VI, Julius-Maximilian University, Am Hubland, 97074 Würzburg, Germany
[b] Center for Applied Energy Research e.V. (CAE), Magdalena-Schoch Straße 3, 97074 Würzburg, Germany

The interplay of host rigidity, dihedral rotation, and polarity with TADF dynamics is illustrated in Fig. 1 (left). The energetic arrangement of $^{1}$CT, $^{3}$CT and $^{3}$LE states is directly targeted by the host polarity, hereby effecting vibronic and spin-orbit coupling. Host rigidity imposes constrains to the conformational freedom along the dihedral angle, thereby indirectly influencing the energetic arrangement and vibronic coupling. This leads to a distribution of relevant TADF rates across individual emitters, reflecting their specific local surroundings. Accessing such distributions would deepen our understanding of host-emitter interactions at the molecular scale, especially regarding the impact of microscopic energetics and dynamics on the macroscopic performance of TADF-based OLEDs.

Ensemble techniques, particularly temperature-dependent transient photoluminescence (PL) studies, conceal the local dependence and distribution of TADF properties.[12, 13] Here, we employ single molecule spectroscopy to directly probe host-specific TADF dynamics at the individual emitter, thereby avoiding information loss, due to ensemble averaging. While single photon techniques have been used to investigate emission mechanisms, excitonic processes and emitter orientation, they have not, to our knowledge, been applied to correlate single molecule TADF emission with host rigidity and polarity.[14-24]

To this end, the TADF emitter 2-[4-(diphenylamino)phenyl]-10,10-dioxide-9H-thioxanthen-9-one (TXO-TPA), which has been proven for its applicability in OLEDs[25, 26] (Fig. 1, left), is doped into the wide band gap hosts poly(methylmethacrylate) (PMMA), bis[2-(diphenylphosphino)phenyl]etheroxide (DPEPO) and 1,4-bis(triphenylsilyl)benzene (UGH-3) (Fig. 1, right). Due to their amorphous, flexible nature, PMMA and DPEPO are expected to allow unrestricted dihedral rotation, whereas the polycrystalline UGH-3 likely restricts molecular motion in TXO-TPA.

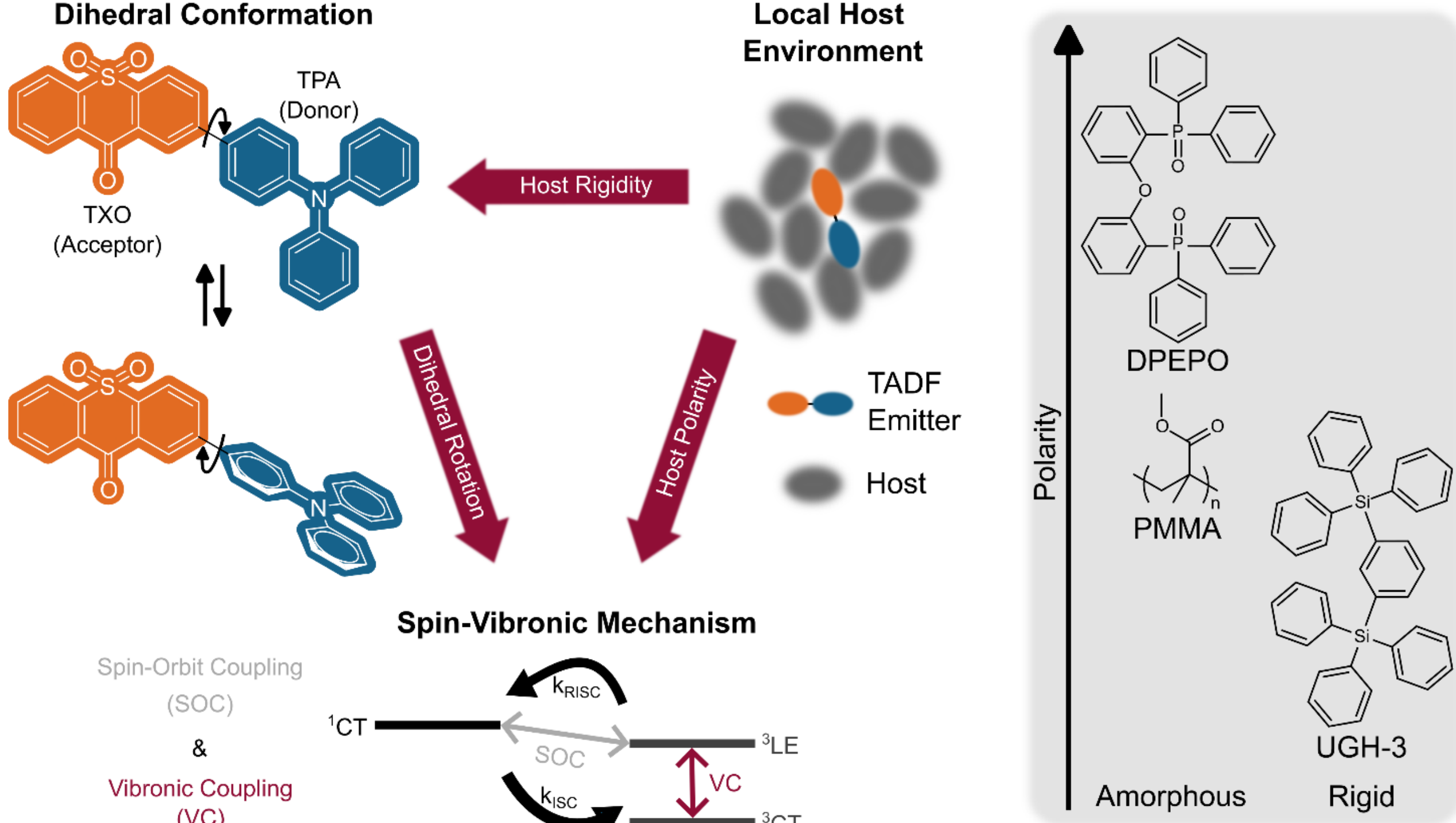

**Fig. 1** Impact of the local host environment on singlet-triplet dynamics of the D-A type TADF emitter TXO-TPA. Red arrows indicate how host polarity, rigidity, and dihedral rotation influence the spin-vibronic mechanism. Single molecule spectroscopy can disclose such local interactions without ensemble averaging. Shown right are the molecular structures of the host materials used in this study: PMMA and DPEPO (both amorphous, with DPEPO highly polar) and UGH-3 (rigid, low polarity).

Additionally, DPEPO provides a highly polar environment due to the phosphine oxide groups, compared to medium polarity PMMA and low polarity UGH-3. Rigidity and polarity effects on the spin-vibronic mechanism are systematically assessed via photon correlation experiments on single molecules of each host-emitter combination. To ensure consistency, intensity histograms, antibunching, and bunching correlations are acquired within the same lifetime cycle, while host-dependent vibronic coupling is probed separately via single molecule PL spectra. All single molecule measurements are carried out at 532 nm cw excitation and 300 µW excitation power (≈ $4{\cdot}10^{-10}$ $W{\cdot}nm^{-2}$) with circularly polarized light. Crucially, we not only reveal host-emitter interactions on the single molecule level but directly link them to ISC/RISC dynamics relevant for the rational selection of host-emitter combinations in OLEDs.

## Results and Discussion

### Spectral Emission

To identify and analyze host effects on the vibrational and conformational freedom of TXO-TPA, crucial for interpreting photon correlation data, we begin our study by discussing single molecule PL spectra. This analysis most effectively illustrates the unique capabilities of single molecule spectroscopy, particularly to researchers less familiar with the technique.

In Fig. 2, representative single molecule spectra from three individual TXO-TPA molecules for each host material, are depicted. For comparison, the corresponding ensemble PL spectra, are shown. In the single molecule studies, a TXO-TPA concentration of $10^{-8}$ wt% results in an average in-plane spacing between individual emitter molecules of at least 1.5 µm, exceeding the confocal excitation diameter by more than a factor of two. In contrast, the 1 wt% emitter concentration used for ensemble PL measurements corresponds to typical TADF study conditions. Self-absorption is negligible at this concentration, given the weak CT transition of TXO-TPA in the visible region.

In ensemble measurements, TXO-TPA features a broad and structureless PL centered at 1.99 eV (623 nm) in PMMA, 1.95 eV (636 nm) in DPEPO and 2.05 eV (605 nm) in UGH-3. The emission peak position generally reflects the average dielectric environment. Polar hosts stabilize the $^1$CT excited state relative to $S_0$, resulting in a bathochromic shift of the emission. Accordingly, DPEPO induces a red shift compared to the nonpolar UGH-3, while PMMA, being moderately polar, yields an intermediate spectral position. The spectral position of the TXO-TPA ensemble PL is in accordance with the polarity sequence of the host materials. The vibronic fine structure is fully averaged out, masking host-specific vibronic coupling.

In contrast, the single molecule PL spectra clearly reveal a vibronic progression, corresponding to 0-0, 0-1, 0-2 and 0-3 transitions between the $^1$CT excited state and $S_0$ ground state. Additional single molecule spectra are shown in SI 1 to SI 3 (Fig. S1 to S3). The quantities extracted below are averaged from at least 10 individual molecules. The ensemble spectrum can be understood as an envelope of many such individual spectra, each shaped by the specific microscopic environment.

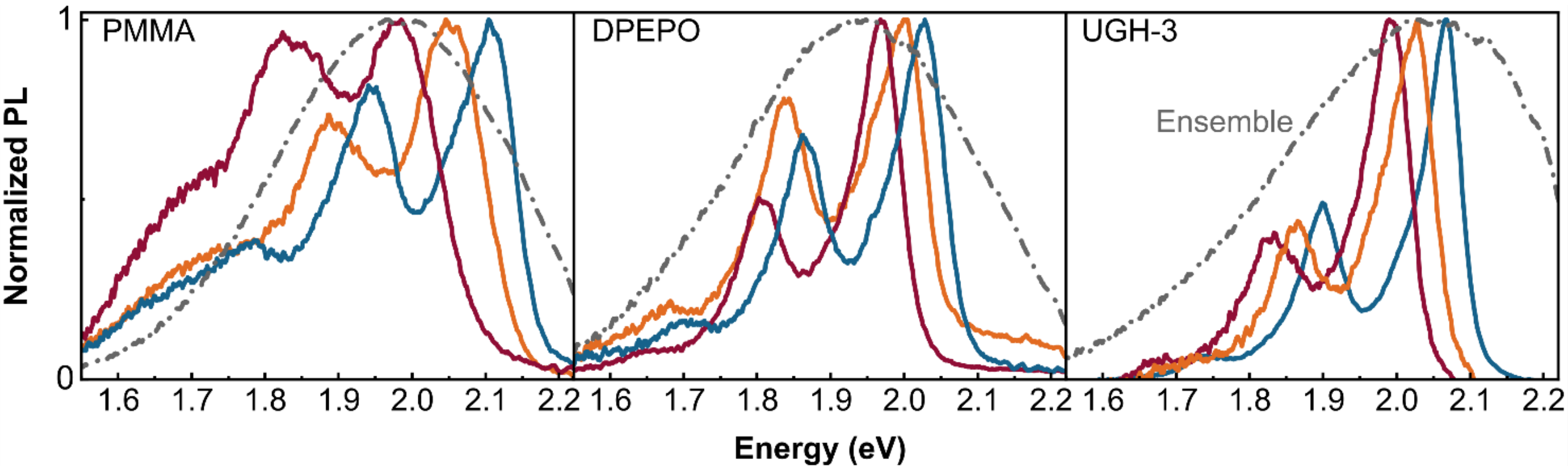

**Fig. 2** Normalized single molecule PL spectra of TXO-TPA in PMMA, DPEPO and UGH-3 (left to right) at ultralow concentrations of $10^{-8}$ wt%. Representative spectra of three individual molecules per host are shown (blue, orange, red solid lines). Ensemble PL spectra at emitter concentrations of 1 wt% are depicted for comparison (dashed-dotted lines).

The observed variation in spectral shape and position across individual TXO-TPA molecules within each host provides direct evidence for the influence of local host-emitter interactions, subtle effects that remain inaccessible in ensemble measurements. The asymmetric peak shapes, deviating from Gaussian or Lorentzian profiles, indicate a superposition of multiple vibrational modes. The 0-0 and 0-1 transitions are separated by 0.16 - 0.17 eV across all hosts, corresponding to the average vibrational mode energy of a single TXO-TPA molecule. Vibronic linewidths differ significantly between hosts. The average full width at half maximum (FWHM) of the 0-0 transition is 0.13 ± 0.01 eV in PMMA, 0.10 ± 0.01 eV in DPEPO and 0.07 ± 0.01 eV in UGH-3. This implies restricted molecular vibrations and dihedral rotation in the rigid UGH-3 host. The mean energetic positions of the 0-0, 0-1 and 0-2 transitions, listed in SI 4 (Table S1), do not fully correlate with host polarity, indicating that host-specific vibronic coupling might affect the spectral position of the ensemble PL. Vibronic coupling is characterized by the Huang-Rhys factor $S$, which quantifies the coupling of the electronic excitation to specific vibronic modes and can be estimated from the intensity ratio of the 0-0 and 0-1 vibrational peak by $I_{0-1}/I_{0-0}$.[27] $S$ is 0.58 ± 0.27 in UGH-3, increasing to 0.66 ± 0.17 in DPEPO and 1.01 ± 0.24 in PMMA, indicating reduced vibronic coupling in the rigid UGH-3 host. When considering a spin-vibronic TADF mechanism, a variation in vibronic coupling may strongly alter the TADF dynamics. As we will later show, the reduced vibronic coupling in UGH-3 does not impair efficient RISC of TXO-TPA molecules being in an energetically favourable dihedral conformation.

**Photon Emission Rate**

The host-specific conformational freedom of TXO-TPA is further investigated by analyzing histograms of the photon emission rate (Fig. 3), derived from intensity time traces recorded during the photon correlation measurements. An intensity time trace binned at 10 ms was converted into a histogram by plotting the photon emission rate against its frequency of occurrence across time bins. Peak broadening or the appearance of multiple peaks suggests that the emission rate varied over time. It is important to note that the chosen binning interval is significantly longer than the triplet lifetime of TXO-TPA (< 100 µs), therefore the data do not resolve singlet-triplet transitions directly. Instead, the histograms reflect the occupation probability of different emissive states.

Fig. 3 shows representative intensity histograms of four individual TXO-TPA molecules in PMMA and DPEPO, and one representative example for UGH-3. Displayed are (multi-)Gaussian fits to the raw data.

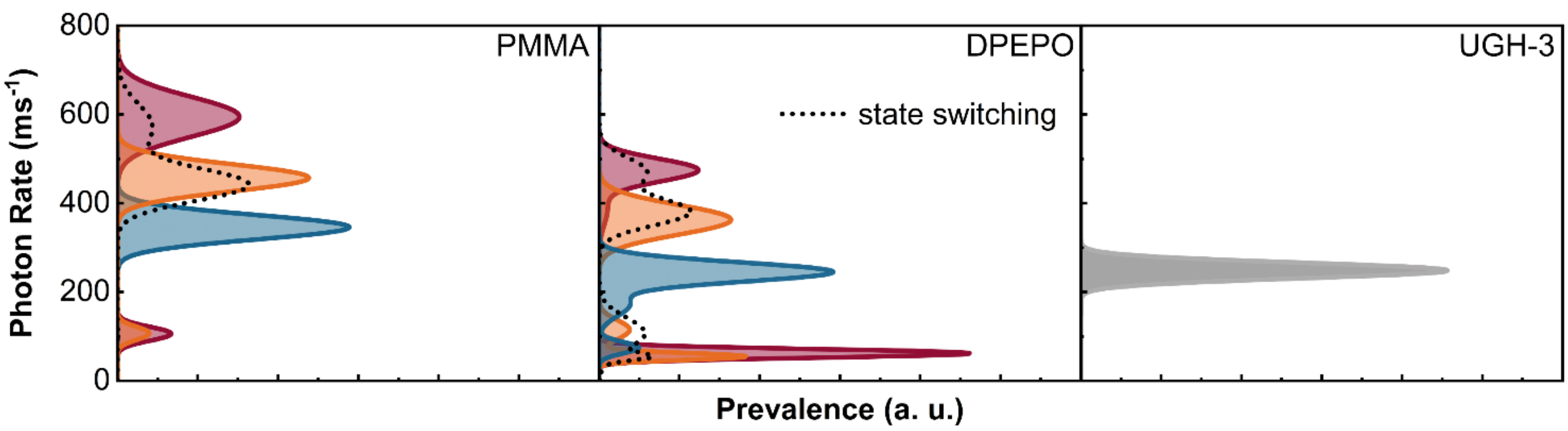

**Fig. 3** Histograms of the photon emission rate of single TXO-TPA molecules in PMMA, DPEPO and UGH-3 (left to right). The histograms are derived from intensity time traces recorded with 10 ms binning. Photon emission rates are plotted against their prevalence. (Multi-) Gaussian fits are shown for clarity, corresponding raw data and fits are provided in SI 6 to SI 8 (Fig. S4 to S6). For PMMA and DPEPO, three representative histograms illustrate characteristic emissive states, along with one example showing switching between states (black dotted line). Peaks below 200 $ms^{-1}$ are assigned to dark states, where the emitter is trapped in the triplet state or irreversibly bleached, and only residual host emission is detected. For UGH-3, one representative histogram reveals a single, narrow emissive state.

The qualitative behaviour captured in these histograms, is representative of the full set of analyzed molecules (SI 6 to SI 8, Fig. S4 to S6). For the amorphous PMMA and DPEPO hosts, we identify three distinct emissive states of TXO-TPA. The mean photon emission rates (with FWHM in parentheses) in PMMA cluster around 600 (48 ± 4), 460 (36 ± 5) and 350 (30 ± 2) $ms^{-1}$, while in DPEPO they are found at 480 (29 ± 5), 370 (33 ± 2) and 250 (22 ± 2) $ms^{-1}$. Individual molecules may occupy one or more of these states, and in some cases (e.g., the black-dotted curves in Fig. 3), dynamic switching between states is evident. Peaks below 200 $ms^{-1}$ correspond to dark states, where the emitter is trapped in the triplet state or irreversibly bleached. Consequently, only residual host emission is detected in those bins. In contrast, TXO-TPA molecules embedded in the rigid host UGH-3 exhibit a single, narrow emissive state (FWHM = 16 ± 2 $ms^{-1}$). The position of this peak varies across molecules (100 - 350 $ms^{-1}$), but no dynamic switching between states is observed, and peak broadening is minimal.

The width of the emissive states mirrors the host-dependent trend in the width of the 0-0 vibronic transition observed in spectral studies. The suppression of emissive state switching in UGH-3 can be attributed to the rigidity of the host matrix, which restricts dihedral fluctuations of the TXO-TPA molecule. In contrast, the conformational freedom in the amorphous PMMA and DPEPO hosts enables switching between different emissive states, likely linked to distinct dihedral conformers. Peak broadening in these hosts is attributed to continuous variation in the dihedral angle around steady-state conformations.

**Single Photon Correlation**

Photon correlation studies at the single-molecule level enable direct access to host-specific singlet-triplet dynamics of the TADF emitter via antibunching and bunching features, as illustrated in Fig. 4 a and b. Upon cw excitation and detection of single photon events by two avalanche photodiodes (APDs) in a Hanbury-Brown Twiss configuration (Fig. 4 a), the second-order correlation function $g^{(2)}(\tau)$ can be derived as:

$$g^{(2)}(\tau) = \frac{\langle I(t) \cdot I(t+\tau) \rangle}{\langle I(t) \rangle^2} \tag{2}$$

This function captures the conditional probability of detecting a photon at time $t + \tau$ given a photon detection at time $t$, represented as the ratio of intensity expectation values. A simulated $g^{(2)}(\tau)$ function for TXO-TPA (Fig. 4 b, top) shows the characteristic features of antibunching and bunching, with corresponding sketches illustrating the mechanisms (Fig. 4 b, bottom). On the ns timescale we observe $g^{(2)}(\tau) \rightarrow 0$ for $\tau \rightarrow 0$ (antibunching), since an excited molecule can only emit one photon at the same time. The width of the antibunching dip provides information on the singlet-state decay of an emitter continuously excited between the $S_0$ and $S_1$ state. Population of $T_1$ via ISC results in dark periods between photon bunches until the $S_1$ state is repopulated via RISC accompanied with delayed fluorescence. The resulting bunching signature ($g^{(2)}(\tau) > 1$) on µs timescales with an exponentially decaying correlation amplitude gives thus access to the ISC/RISC dynamics of the emitter.

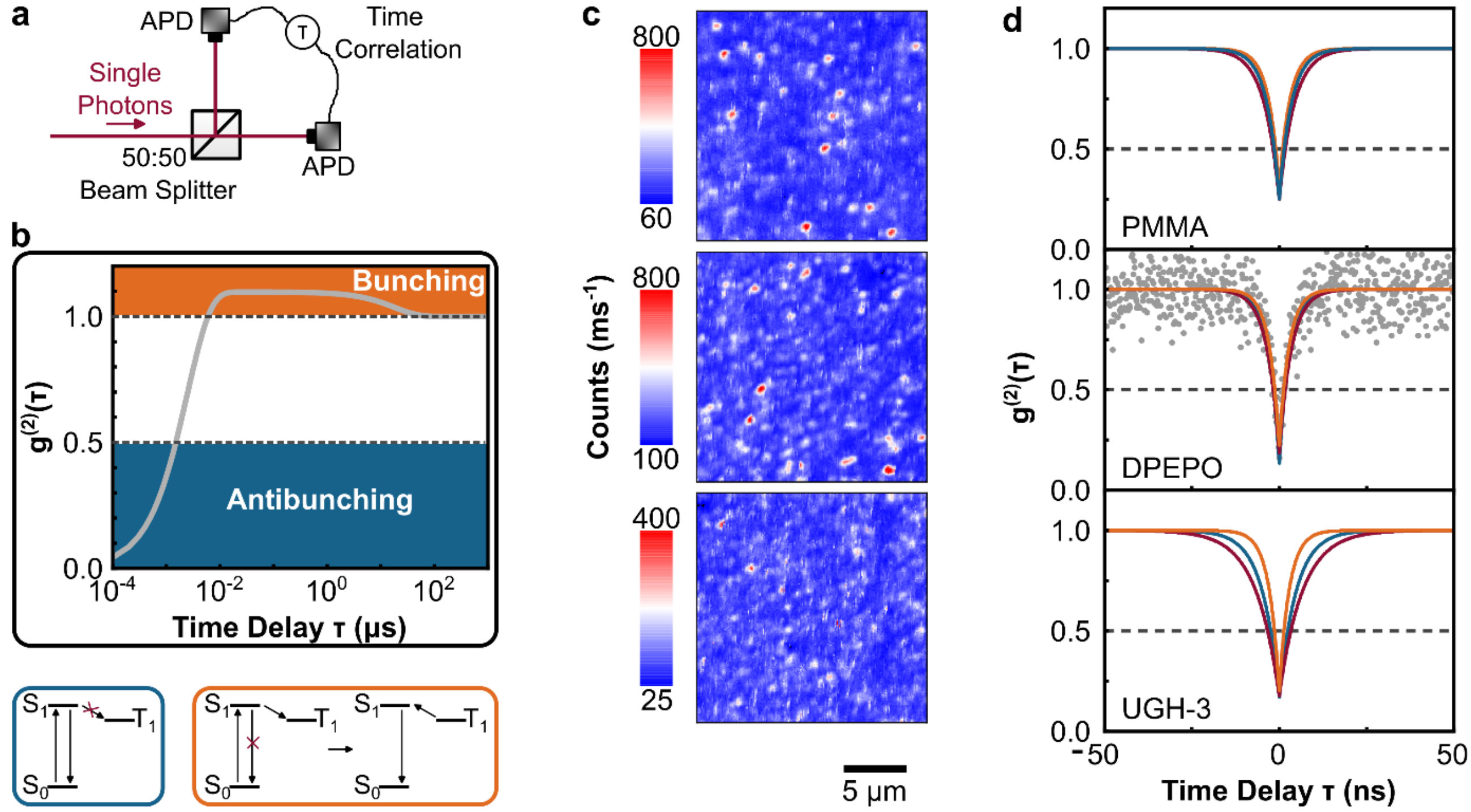

**Fig. 4** Principle of photon correlation measurement, PL maps and antibunching curves. a) Schematic of the Hanbury Brown-Twiss setup with a 50:50 beam splitter and two avalanche photodiodes (APDs). b) Simulated $g^{(2)}(\tau)$ correlation function for TXO-TPA, with excitation-emission cycles for antibunching and bunching sketched in the bottom. c) PL intensity maps of TXO-TPA dispersed at the single molecule level into PMMA, DPEPO and UGH-3 (top to bottom). Color scale minimums are adjusted to the background emission of each host. The maps cover a thin film area of 20 µm². d) Corresponding $g^{(2)}(\tau)$ functions for three individual TXO-TPA molecules per host in the time delay range -50 to 50 ns. Mono exponential fits to the background corrected data are shown and raw data are included for one exemplary molecule in DPEPO. The full set of measured molecules including fits is provided in SI 6 to SI 8 (Fig. S4 to S6). Antibunching with $g^{(2)}(0) < 0.5$ confirms single photon emission in all host materials.

We begin with an analysis of the single photon emission of TXO-TPA dispersed in PMMA, DPEPO and UGH-3. In Fig. 4 c characteristic PL intensity maps are depicted, where isolated and homogenously distributed bright spots reflect emission from individual or few TXO-TPA molecules. The spot size is limited by the confocal resolution of ~ 500 nm. At 300 µW excitation power (≈ $4\cdot10^{-10}$ W nm$^{-2}$), photon count rates range from 100 to 800 ms$^{-1}$. Details on the power dependence of the photon emission rate are given in SI 9 (Fig. S7). The chosen excitation power is a trade-off to excite TXO-TPA near saturation and to minimize photobleaching during correlation measurements.

The definitive proof of single molecule emission is provided by $g^{(2)}(0) < 0.5$, with $g^{(2)}(0) = 1 - 1/N$ indicating the presence of $N$ emitter molecules in the focal volume.[28] Fig. 4 d presents background-corrected antibunching curves for three single TXO-TPA molecules in each host. Mono-exponential fits to the background corrected data are plotted. The full raw data set of all measured molecules and corresponding fits are shown in SI 6 to SI 8 (Fig. S4 to S6). All show clear antibunching dips below 0.5 at $\tau = 0$, confirming single photon emission.

Following Hu et al.[29], the antibunching lifetime $\tau_a$ is given by:

$$g^{(2)}(\tau) = 1 - C \cdot e^{-(k_{abs}+k_r+k_{nr})\cdot\tau} = 1 - C \cdot e^{-\frac{\tau}{\tau_a}} \qquad (3)$$

Here, $C$ is the antibunching amplitude, and $k_{abs}$, $k_r$ and $k_{nr}$ are the absorption, radiative, and non-radiative decay rates, respectively. $k_{abs}$ depends on the excitation power and the absorption cross-section of the molecule, which leads to a deviation of $\tau_a$ from ensemble fluorescence lifetimes. Since excitation was done with circularly polarized cw light and the absorption cross-section is assumed constant, variations in $\tau_a$ mainly reflect changes in $k_r$ and $k_{nr}$. These are expected to be influenced by the local host environment. Comparing the host materials, UGH-3 yields the broadest spread in antibunching lifetimes across single molecules (2.9 - 6.7 ns), in contrast to narrower ranges in PMMA (2.7 - 4.5 ns) and DPEPO (2.8 - 3.5 ns). We analyze at least 10 single molecules for each host. The lower limit is nearly identical for all hosts. We attribute the increased spread in UGH-3 to its rigidity, which restricts conformational reorganization, impacting both radiative and non-radiative decay of the $^1$CT state. This trend mirrors the ensemble prompt fluorescence decay in UGH-3, exhibiting distinct distributions in the fluorescence lifetime (SI 10, Fig. S8). Polarity, by contrast, shows minimal or no effect on the antibunching lifetimes, as also evident from complementary ensemble measurements. Notably, pronounced bunching, as vide infra observed for a few emitters in UGH-3, influences the antibunching width.

To elucidate the host-dependent singlet-triplet dynamics at the single molecule level, we analyzed photon bunching on intermediate timescales within the same lifetime cycle of individual TXO-TPA molecules. Fig. 5 shows bunching measurements from three representative single molecules for each host-guest combination, capturing the variation observed across the full data set (SI 6 to SI 8, Fig. S4 to S6). Bunching is characterized by positive correlations ($g^{(2)}(\tau) > 1.0$). The bunching decay between 0.1 µs and 1000 µs was fitted using the exponential function:

$$g^{(2)}(\tau) = 1 + A \cdot e^{-\frac{\tau}{\tau_b}} \qquad (4)$$

Here the bunching amplitude $A$ serves as a metric for the maximum offset from uncorrelated emission, and the decay constant $\tau_b$ quantifies the characteristic bunching duration. The values presented in the following are derived from fits to at least 10 individual molecules (SI 6 to SI 8, Fig. S4 to S6).

The extracted parameters disclose two host-dependent trends. First, the rigid UGH-3 host displays a binary distribution in bunching amplitude: most measured TXO-TPA molecules exhibit negligible bunching ($A < 0.01$), while an insignificant number showed pronounced bunching with $A > 0.4$. Notably, strong bunching in UGH-3 correlates with broader antibunching widths. In contrast, the amorphous hosts PMMA and DPEPO show mild but consistent bunching for all molecules, with amplitude distributions ranging from 0.01 - 0.13 (PMMA) and 0.01 - 0.11 (DPEPO). Second, the bunching duration $\tau_b$ is related to host polarity with 11 µs for UGH-3, 16 ± 5 µs for PMMA, and 30 ± 9 µs for DPEPO. These host-specific trends at the single-molecule level are hidden in transient PL data of emitter ensembles (SI 10, Fig. S9).

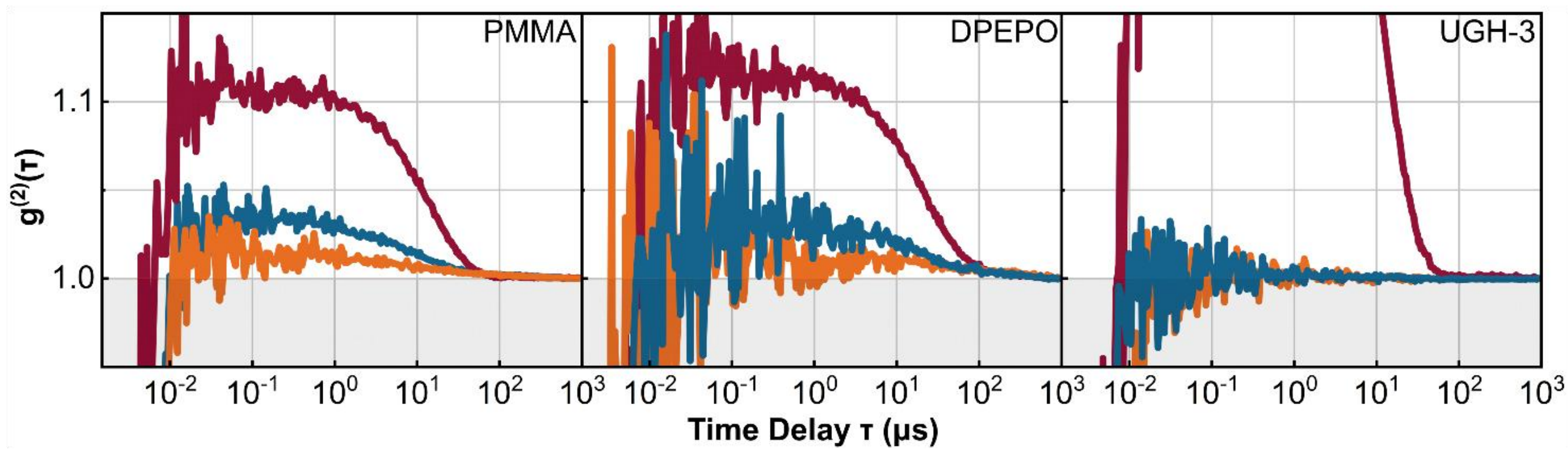

Fig. 5 $g^{(2)}(\tau)$ correlation functions recorded over time delays up to 1000 µs. Shown are raw data of three individual TXO-TPA molecules for each host. The full data set of measured molecules together with exponential fits are displayed in the SI 6 to SI 8 (Fig. S4 to S6). Bunching is evident where $g^{(2)}(\tau) > 1.0$ on intermediate time scales.

To derive concrete host effects on the bunching amplitude distribution and bunching duration, we modelled the correlation function of TXO-TPA based on a set of differential equations representing a generic three-level system with associated transition rate constants.

We attribute the observed differences in bunching duration between the high polarity host DPEPO, the medium polarity host PMMA, and the low polarity host UGH-3 to the dielectric properties of the respective hosts. The simulation in Fig. 6a highlights how bunching duration depends on the absolute magnitudes of $k_{RISC}$ and $k_{ISC}$. We infer that the enhanced bunching duration observed for individual TXO-TPA molecules in DPEPO arises from a polarity-induced reshaping of the energetic landscape involving the $^1$CT, $^3$LE, and $^3$CT states. A polarity-dependent mechanism is illustrated in Fig. 6c. In low to medium polarity environments, the energetic proximity of the $^1$CT and $^3$LE states facilitates strong spin-orbit coupling, leading to fast ISC/RISC dynamics. Conversely, in a polar host like DPEPO, stabilization of the CT states lowers their energies relative to the $^3$LE state, reducing spin-orbit coupling due to an increase in the energy gap between $^1$CT and $^3$LE. Consequently, ISC and RISC are decelerated. From the simulations, we deduce that the $k_{RISC}$ and $k_{ISC}$ values of TXO-TPA molecules in DPEPO are approximately halved compared to those in PMMA. In general, the variation in bunching durations over individual emitters reflects local host polarity differences, which are pronounced in DPEPO.

The simulation in Fig. 6b reveals that the bunching amplitude decreases as $k_{RISC}$ approaches $k_{ISC}$, eventually eliminating the bunching signature. This condition may occur when the dihedral conformation of TXO-TPA results in a beneficial arrangement of $^1$CT, $^3$LE, and $^3$CT states, enabling strong spin-orbit and vibronic coupling. Together with steady-state spectroscopic data and intensity histograms discussed earlier, we identify dihedral conformation-driven RISC dynamics as the primary factor determining the bunching amplitude of individual emitters, with the mechanism illustrated in Fig 6d. In the amorphous hosts PMMA and DPEPO, conformational flexibility enables the dihedral angle to sample a range of geometries near equilibrium positions. As a result, TXO-TPA exhibits a broad distribution of $k_{RISC}$ values ranging from $10^4$ s$^{-1}$ to $10^6$ s$^{-1}$, which we refer to a variation in vibronic coupling due to conformation driven energetic spacing between $^3$LE and $^3$CT. We emphasize that while these rate constants are illustrative and not absolute, the variation in bunching amplitude is mainly governed by the ratio $k_{RISC}/k_{ISC}$. In the rigid host UGH-3, steric constraints enforce TXO-TPA into a dominant dihedral conformation, associated with fast RISC ($k_{RISC} \approx 10^6$ s$^{-1}$). In this dihedral conformation, both spin-orbit coupling and vibronic coupling are enhanced due to energetically close $^1$CT, $^3$CT and $^3$LE states.

Our findings are consistent with the hypothesis proposed by Miranda-Salinas et al.[7], who suggested a conformation-driven RISC mechanism for the TADF emitter TpAT-tFFO in UGH-3. However, unlike their study, our single-molecule analysis provides direct insight at the molecular scale, supporting the critical role of conformational rigidity. Notably, despite the suppression of vibronic coupling and vibrational motion - visible in both PL spectra and intensity histograms - the fast RISC dynamics in UGH-3 appear to be primarily governed by conformational energy alignment, rather than by vibrational freedom. Contrary to expectations, RISC is more efficient in the rigid host despite reduced molecular motion, which is often considered crucial for $^3$CT to $^3$LE transitions.

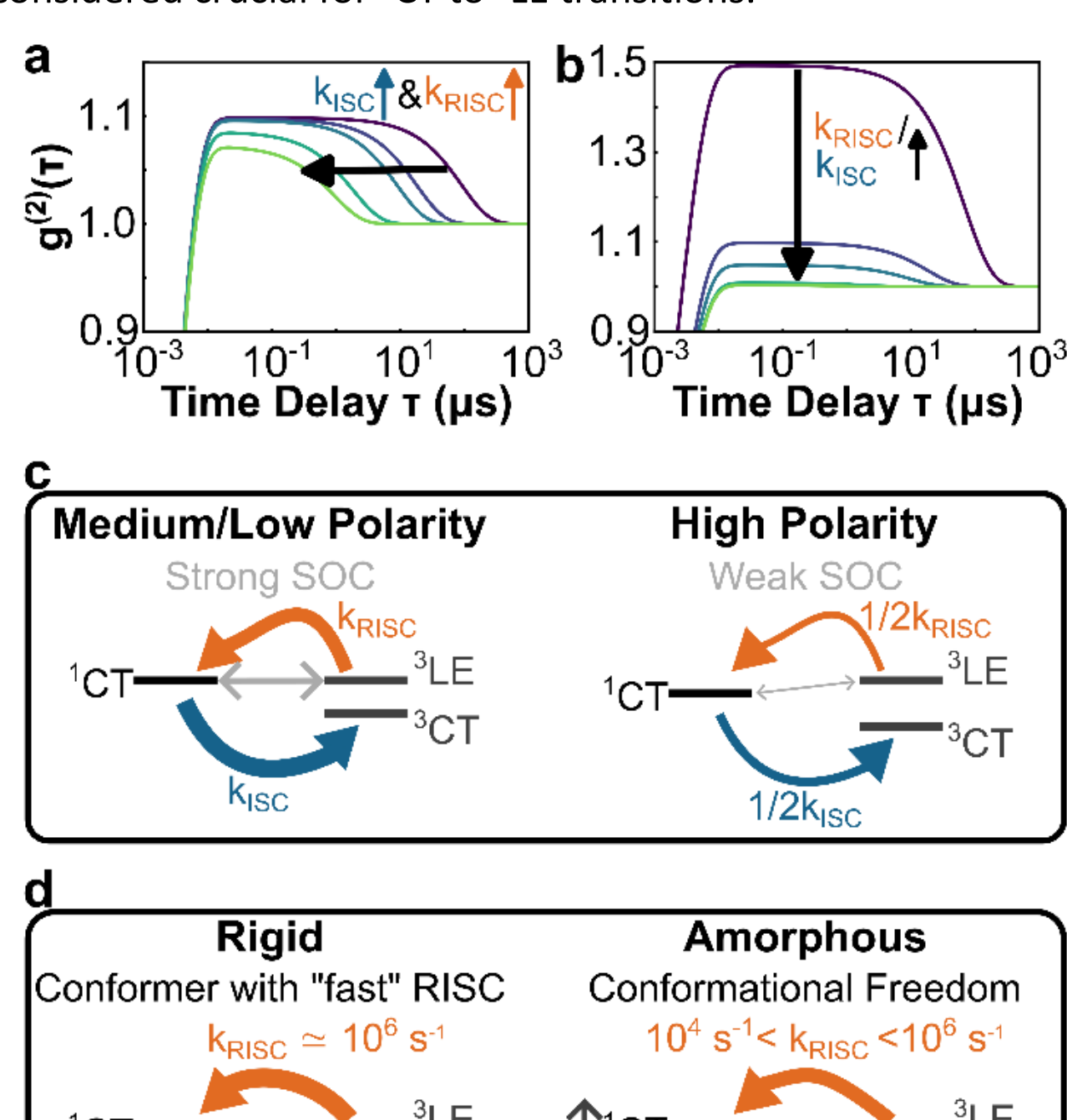

Fig. 6 Simulation of correlation functions and proposed host-dependent mechanisms. a) Simulated evolution of the bunching duration as a function of increasing $k_{RISC}$ and $k_{ISC}$, with both rate constants varied simultaneously from $10^4$ s$^{-1}$/$10^5$ s$^{-1}$ to $10^6$ s$^{-1}$/$10^7$ s$^{-1}$. b) Simulated reduction in bunching amplitude as result of increasing $k_{RISC}$ relative to $k_{ISC}$, with the ratio $k_{RISC}/k_{ISC}$ varied from 0.01 to 1. c) Proposed polarity-related host mechanism affecting the bunching duration. In low to medium polarity hosts (UGH-3, PMMA), the energetic proximity of the $^3$LE and $^1$CT states enhances spin-orbit coupling (SOC) resulting in fast ISC/RISC dynamics. In contrast, in the high polarity host DPEPO, stabilization of $^1$CT relative to $^3$LE reduces SOC and slows ISC/RISC processes. Variations in bunching duration between individual emitters reflect local host polarity differences. d) Proposed rigidity-related host mechanism affecting the bunching amplitude. In the rigid UGH-3 host, most TXO-TPA molecules are enforced into a dihedral geometry with energetically close $^1$CT, $^3$CT and $^3$LE states, enhancing both SOC and vibronic coupling (VC) and leading to efficient RISC. In contrast, amorphous hosts (PMMA, DPEPO) allow conformational freedom, resulting in varying energies of the $^1$CT and $^3$CT states and thus broad distributions in VC between $^3$LE and $^3$CT. Consequently, $k_{RISC}$ varies over two orders of magnitude across individual emitter molecules.

## Conclusions

We investigated the single molecule emission of the D-A type TADF emitter TXO-TPA embedded at ultralow concentrations in technologically relevant host materials, revealing local environmental effects on emitter dynamics. Using steady-state spectroscopy, intensity histograms, and photon correlation measurements, we directly accessed host-dependent TADF dynamics without the averaging effects inherent to ensemble measurements.

Single-molecule PL spectra show that molecular vibrations and vibronic coupling are significantly suppressed in the rigid host UGH-3, evidenced by narrower spectral features and lower Huang-Rhys factors. Despite restricted molecular motion, most TXO-TPA molecules in UGH-3 adopt a dihedral conformation supporting fast RISC ($k_{RISC} \approx 10^6\ s^{-1}$), as indicated by the absence of photon bunching and supported by simulations. In contrast, the amorphous hosts PMMA and DPEPO permit conformational flexibility, resulting in distinct emissive states and broad distributions in $k_{RISC}$ ($10^4\ s^{-1}$ to $10^6\ s^{-1}$), reflected in the bunching amplitude variation. Our findings demonstrate how host rigidity affects the energetic arrangement of the $^1$CT, $^3$CT and $^3$LE excited states within the emitter.

Furthermore, we observe a strong influence of local host polarity on the bunching duration. In the polar DPEPO host, stabilization of the $^1$CT state reduces spin-orbit coupling and slows ISC/RISC dynamics, leading to shorter bunching durations compared to PMMA.

In summary, this study reveals how host polarity and rigidity shape TADF dynamics at the molecular level, offering key insights for the rational design of optimized host-emitter combinations in OLED applications.

## Experimental

**Preparation of Thin Films.** Thin films of TXO-TPA dispersed in various host materials were prepared by spin-coating under inert conditions ($N_2$ glovebox, $O_2 < 0.1$ ppm, $H_2O < 0.1$ ppm). PMMA ($M_w = 300{,}000\ g\cdot mol^{-1}$), UGH-3, DPEPO (all Sigma Aldrich), and TXO-TPA (Lumtec) were used as received at sublimed grade. Microscope cover slides (Karl Hecht, 170 µm thickness) were cleaned via sequential ultrasonication in double-distilled water with mucasol® detergent, distilled water, acetone, and isopropanol, followed by nitrogen drying. Mixtures of the host materials and a highly diluted weight fraction of TXO-TPA ($10^{-8}$ wt%) were dissolved in DCM (Sigma Aldrich, spectroscopy grade, $10\ mg\cdot mL^{-1}$) and spin-coated at 2000 rpm for 60 s to yield films of 100 - 150 nm thickness. Afterward, samples were thermally treated at 70 °C for 45 min to remove residual solvent. Parts of the films were covered with aluminum (Al, 99.9999%, Chempur) as an oxygen getter. Al (120 nm) was thermally evaporated in a glovebox-integrated vacuum chamber (base pressure: $1 \cdot 10^{-6}$ mbar; rate: 2 - 4 $Å\cdot s^{-1}$) using tungsten boats, with thickness monitored via a quartz crystal microbalance. Finally, the samples were encapsulated with cover glass and epoxy resin (Loctite EA9492). The Al-covered regions remained within the sealed volume to ensure oxygen exclusion during single-molecule measurements.

**Single Molecule Spectroscopy.** Single-molecule PL and photon-correlation measurements were conducted on a home-built confocal microscopy setup. A 532 nm cw laser (CNI) filtered through a spatial mode filter (Thorlabs) was used for excitation. Circularly polarized light was achieved using a quarter- and half-wave plate to minimize polarization effects. Excitation power was adjusted with a neutral density wheel and monitored with a calibrated powermeter (Thorlabs S130C). Samples were scanned via a piezo stage (nPoint, 200 x 200 $\mu m^2$ range). A high NA oil-immersion objective (Olympus, 100x, NA = 1.49) provided ~ 500 nm lateral resolution. Emission was collected in epifluorescence configuration using a dichroic mirror, razor-edge filter, and 550 nm long-pass filter. A reflective two-lens system and 75 µm pinhole suppressed background light. Spectra were recorded using a spectrometer (Princeton Instruments, Acton SP2300) coupled to a CCD camera (Pixis 400B). Photon correlation was performed using two single-photon avalanche diodes (Excelitas SPCM-AQRH-14, QE 65 % at 650 nm, dark count < 100 cps) in a Hanbury Brown-Twiss setup with a 50:50 beam splitter (Thorlabs). Correlation functions $g^{(2)}(\tau)$ were obtained using a hardware correlator (Becker & Hickl, DPC-230, 350 ps resolution, limited by detector jitter). Histograms of the photon emission rate were captured from intensity time traces with a binning time of 10 ms. Antibunching and bunching data as well as histograms of the photon emission rate were recorded within the same measurement cycle for each single molecule. Background correction was applied to antibunching curves using intensity measured near the molecule (see SI5, Eq. SI2). The spectral data were post-processed by a moving average filter (10 points) and a Jakobi transformation from the wavelength to the energy regime. All measurements were conducted at 300 µW cw excitation, corresponding to ~ $4 \cdot 10^{-10}\ W\cdot nm^{-2}$.

**Simulation and Modelling of Correlation Functions.** The following set of differential equations describing the time-dependent change in population of the $S_0$ ($p_1$), $S_1$ ($p_2$) and $T_1$ ($p_3$) states was analytically solved for $p_2(\tau)$ with $\tau$ being the time delay and the rate constants $k$ describing transitions in the generic three-state model of a TADF emitter.

$$(I)\ \dot{p}_1 = -k_{12}p_1 + k_{21}p_2 + k_{31}p_3$$

$$(II)\ \dot{p}_2 = k_{12}p_1 - (k_{21} + k_{23})p_2 + k_{32}p_3$$

$$(III)\ \dot{p}_3 = k_{23}p_2 - k_{31}p_3 - k_{32}p_3$$

Following Hecht et al.[30] the analytical $g^{(2)}(\tau)$ function was derived as $g^{(2)}(\tau) = p_2(\tau)/p_2(\infty)$. Base rate constants were obtained from ensemble measurements. Simulations of bunching characteristics were performed using Python.

## Author contributions

Conceptualization (BE, JP), data curation (BE), formal analysis (BE, TF), funding acquisition (JP), investigation (BE; TK, TF), project administration (BE, JP), resources (JP), supervision (JP, BE), validation (BE, TK, TF), visualization (BE), writing-original draft (BE), writing - review & editing (JP, BE).

## Conflicts of interest

There are no conflicts to declare.

## Data availability

The data supporting this article have been included as part of the Supplementary Information.

## Acknowledgements

German Research Foundation (project PF385/12-1)
Bavarian State Ministry of Science, Research, and the Arts (Collaborative Research Network "Solar Technolgies Go Hybrid")

# Supporting Information

## Disclosing the Impact of Local Host Effects on TADF Dynamics

Björn Ewald,*[a] Theodor Kaiser,[a] Thomas Fleischmann[a] and Jens Pflaum*[a,b]

[a] *Experimental Physics VI, Julius-Maximilian University, Am Hubland, 97074 Würzburg, Germany*

[b] *Center for Applied Energy Research e.V. (CAE), Magdalena-Schoch Straße 3, 97074 Würzburg, Germany*

### Corresponding Authors

Björn Ewald - Experimental Physics VI, Julius-Maximilian University, Am Hubland, 97074 Würzburg, Germany
https://orcid.org/0009-0000-7078-9252;
Email: bjoern.ewald@uni-wuerzburg.de

Jens Pflaum - Experimental Physics VI, Julius-Maximilian University, Am Hubland, 97074 Würzburg, Germany - Center for Applied Energy Research e.V. (CAE), Magdalena-Schoch Straße 3, 97074 Würzburg, Germany
https://orcid.org/0000-0001-5326-8244
Email: jpflaum@physik.uni-wuerzburg.de

## SI 1: Single Molecule Spectra of TXO-TPA in PMMA

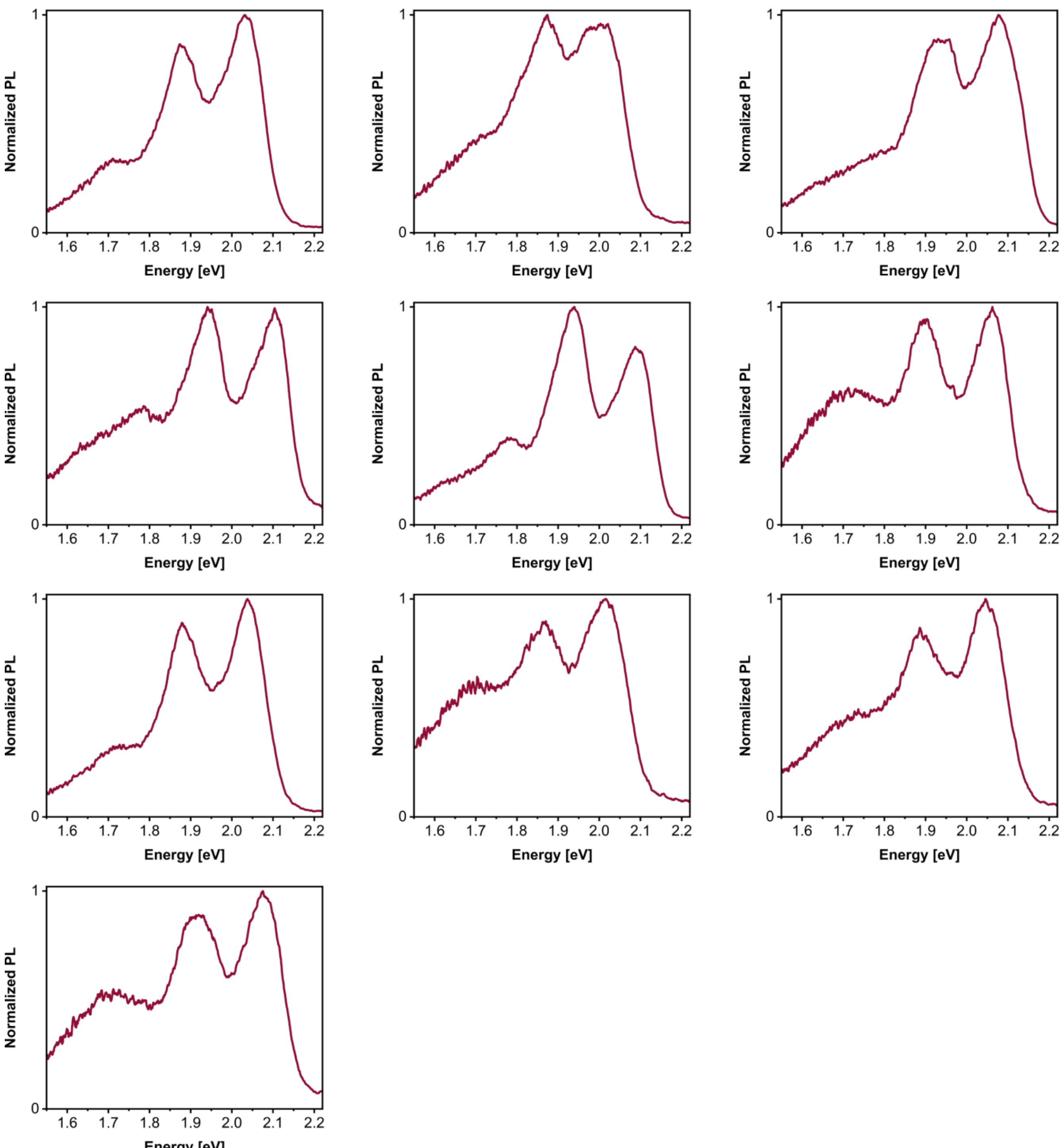

**Figure S1.** Normalized single molecule PL spectra of TXO-TPA dispersed in PMMA at ultralow concentrations of $10^{-8}$ wt%. Representative spectra of individual molecules are shown to illustrate the distribution of spectral properties. The spectral data were post-processed by a moving average filter and a Jakobi transformation from the wavelength to the energy regime.

## SI 2: Single Molecule Spectra of TXO-TPA in DPEPO

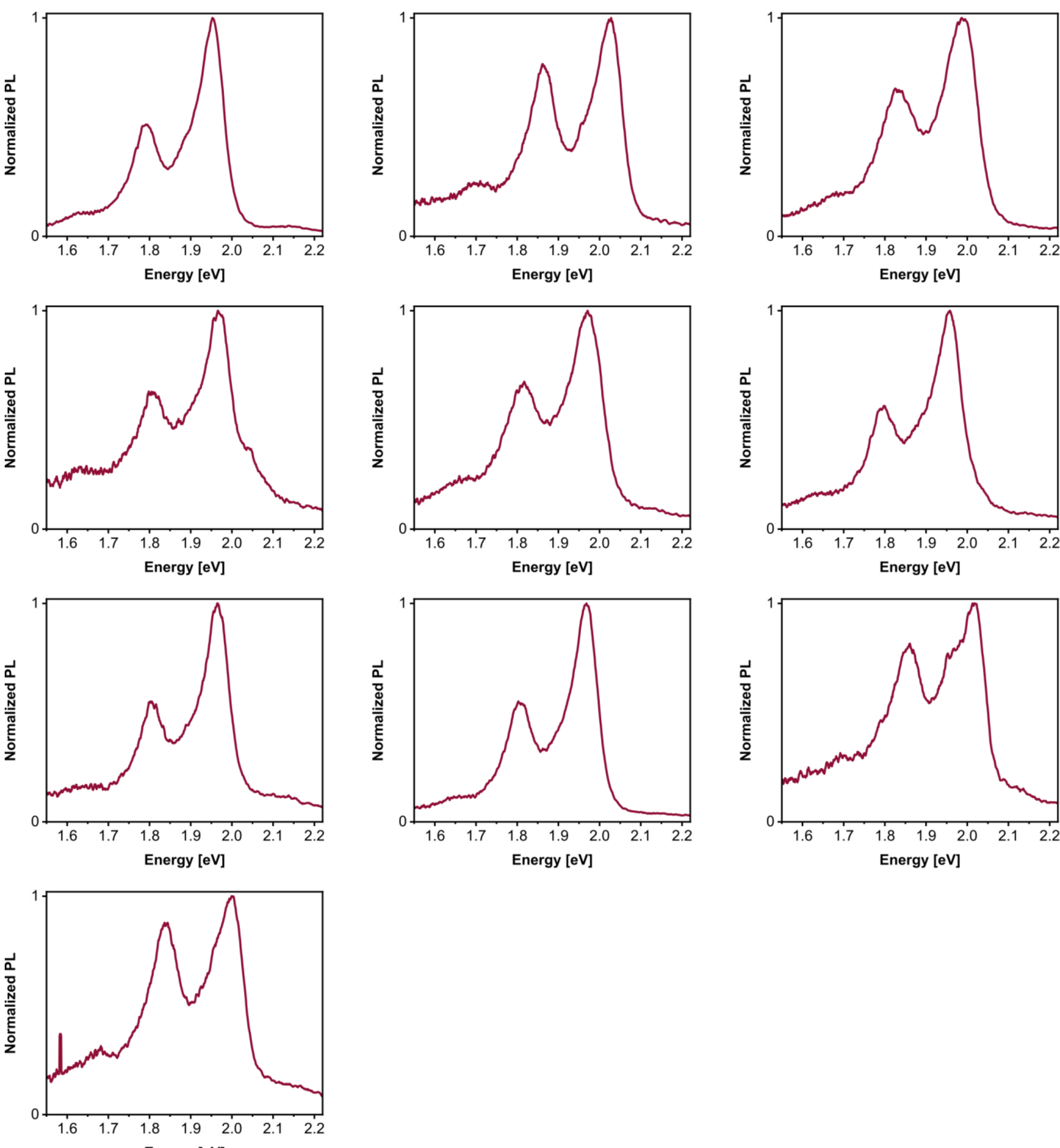

**Figure S2.** Normalized single molecule PL spectra of TXO-TPA dispersed in DPEPO at ultralow concentrations of $10^{-8}$ wt%. Representative spectra of individual molecules are shown to illustrate the distribution of spectral properties. The spectral data were post-processed by a moving average filter and a Jakobi transformation from the wavelength to the energy regime.

## SI 3: Single Molecule Spectra of TXO-TPA in UGH-3

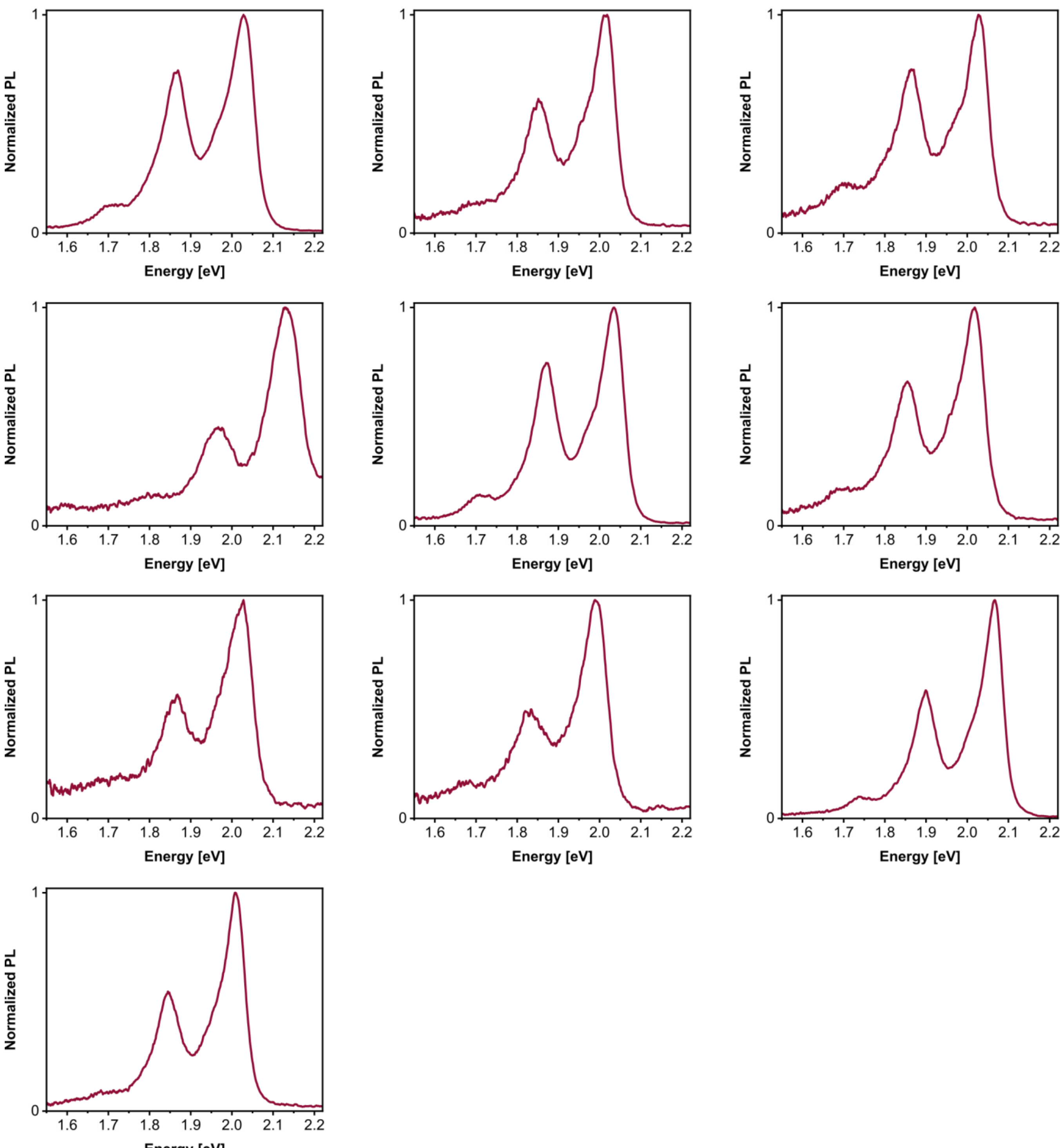

**Figure S3.** Normalized single molecule PL spectra of TXO-TPA dispersed in UGH-3 at ultralow concentrations of $10^{-8}$ wt%. Representative spectra of individual molecules are shown to illustrate the distribution of spectral properties. The spectral data were post-processed by a moving average filter and a Jakobi transformation from the wavelength to the energy regime.

## SI 4: Spectral Properties of Single TXO-TPA Molecules

The spectral properties of single TXO-TPA molecules in PMMA, DPEPO and UGH-3 are listed in Table S1. Given are the energetic positions of the 0-0, 0-1 and 0-2 transitions, as well as the Huang-Rhys parameter *S*. *S* was derived from the integral intensity ratio of the 0-0 and 0-1 transition:

$$S = \frac{I_{0-1}}{I_{0-0}} \qquad \text{(SI1)}$$

**Table S1.** Spectral properties of single TXO-TPA molecules in the three host materials PMMA, DPEPO and UGH-3. The data were derived from a representative set of molecular entities (at least 10 single molecules). Given are the average energetic positions of the vibronic main transitions 0-0, 0-1 and 0-2, as well as the Huang-Rhys parameter *S* calculated from ratio of the integral intensities of the 0-0 and 0-1 transition.

| Host Material | $E_{0\text{-}0}$ (eV) | $E_{0\text{-}1}$ (eV) | $E_{0\text{-}2}$ (eV) | $S$ |
|---|---|---|---|---|
| PMMA | 2.047 ± 0.037 | 1.906 ± 0.031 | 1.768 ± 0.029 | 1.01 ± 0.24 |
| DPEPO | 1.975 ± 0.024 | 1.826 ± 0.024 | 1.670 ± 0.027 | 0.66 ± 0.17 |
| UGH-3 | 2.029 ± 0.040 | 1.871 ± 0.038 | 1.735 ± 0.039 | 0.58 ± 0.27 |

## SI 5: Processing of Single Photon Correlation and Histogram Data

To account for uncorrelated background intensity $I_B$ affecting the antibunching measurements, the corrected $g^{(2)}(\tau)$ correlation function was derived from the uncorrected $g_{uc}^{(2)}(\tau)$ function by the following equation:

$$g^{(2)}(\tau) = \frac{g_{uc}^{(2)}(\tau) - 1 + \left(\frac{I_M(t)}{I_M(t)+I_B}\right)^2}{\left(\frac{I_M(t)}{I_M(t)+I_B}\right)^2} \quad \text{(SI2)}$$

The background intensity $I_B$ was measured beside the single molecule spot and $I_M(t)$ corresponds to the time-dependent intensity at the single molecule location. The corrected antibunching data, were fitted with the following mono exponential function:

$$g^{(2)}(\tau) = 1 - C \cdot e^{-\frac{\tau}{\tau_a}} \quad \text{(SI3)}$$

Here $\tau_a$ is the effective antibunching lifetime and $C$ is the magnitude of the antibunching at zero delay. As pointed out in the main part $\tau_a$ is not equal to the fluorescence lifetime in ensemble experiments. Only antibunching characteristics with $g^{(2)}(0) < 0.5$ were considered for the analysis. The two branches of the antibunching curves were fitted with a combined function.

The bunching data, were utilized without post-processing. Bunching signatures were fitted with the following exponential decay function between 0.1 and 1000 µs:

$$g^{(2)}(\tau) = 1 + A \cdot e^{-\frac{\tau}{\tau_b}} \quad \text{(SI4)}$$

Here $A$ is the bunching amplitude and $\tau_b$ the decay constant quantifying the characteristic bunching duration.

Histogram data were fitted with a (multi-)gaussian distribution function:

$$Prevalence = A \cdot e^{-\frac{(I-M)}{2\sigma^2}} \quad \text{(SI5)}$$

Here $A$ is a scaled amplitude, $I$ is the photon emission rate, $M$ is the mean of the distribution, and $\sigma$ is the standard deviation of the distribution.

# SI 6: Photon Correlation and Histogram Data of TXO-TPA in PMMA

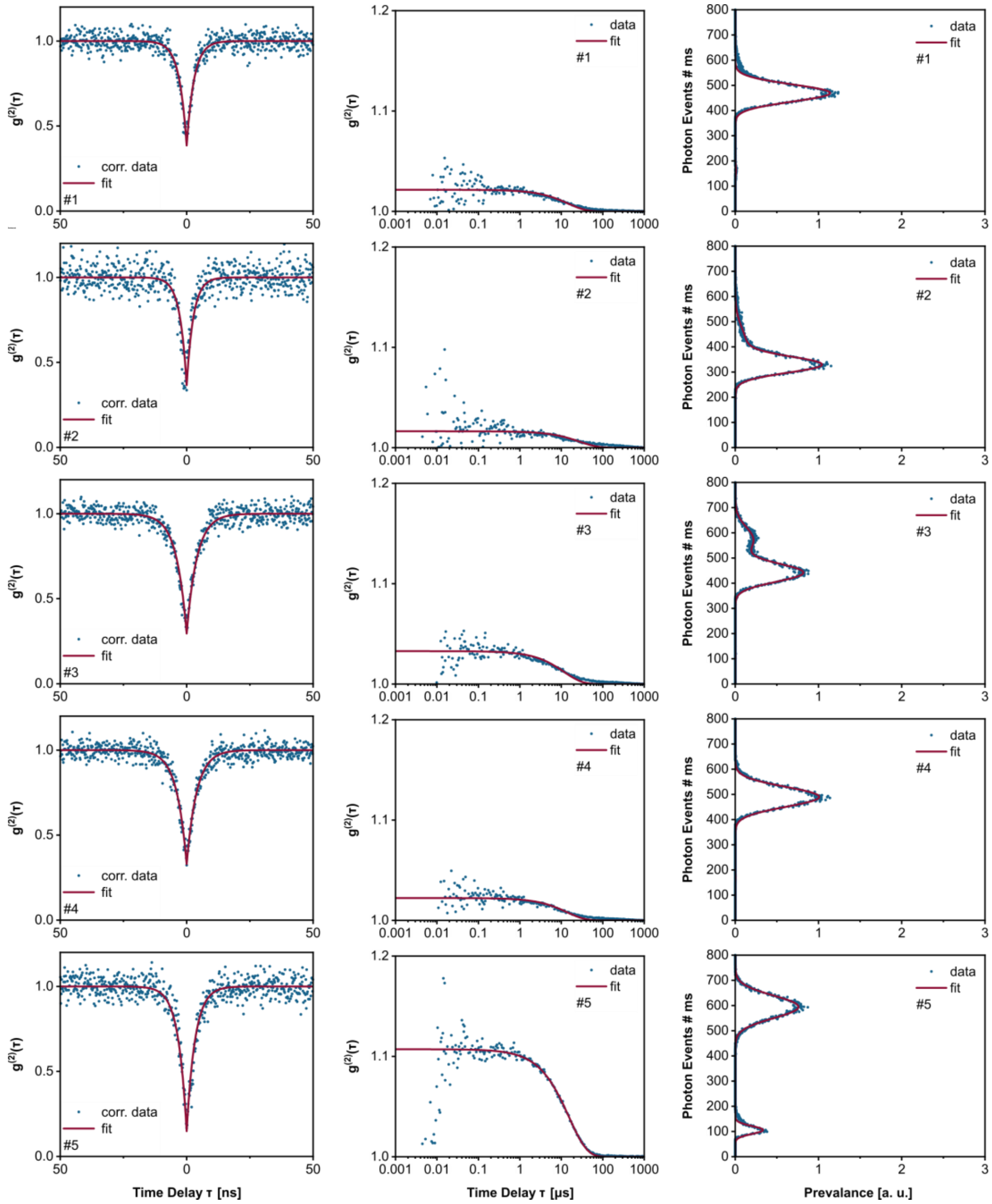

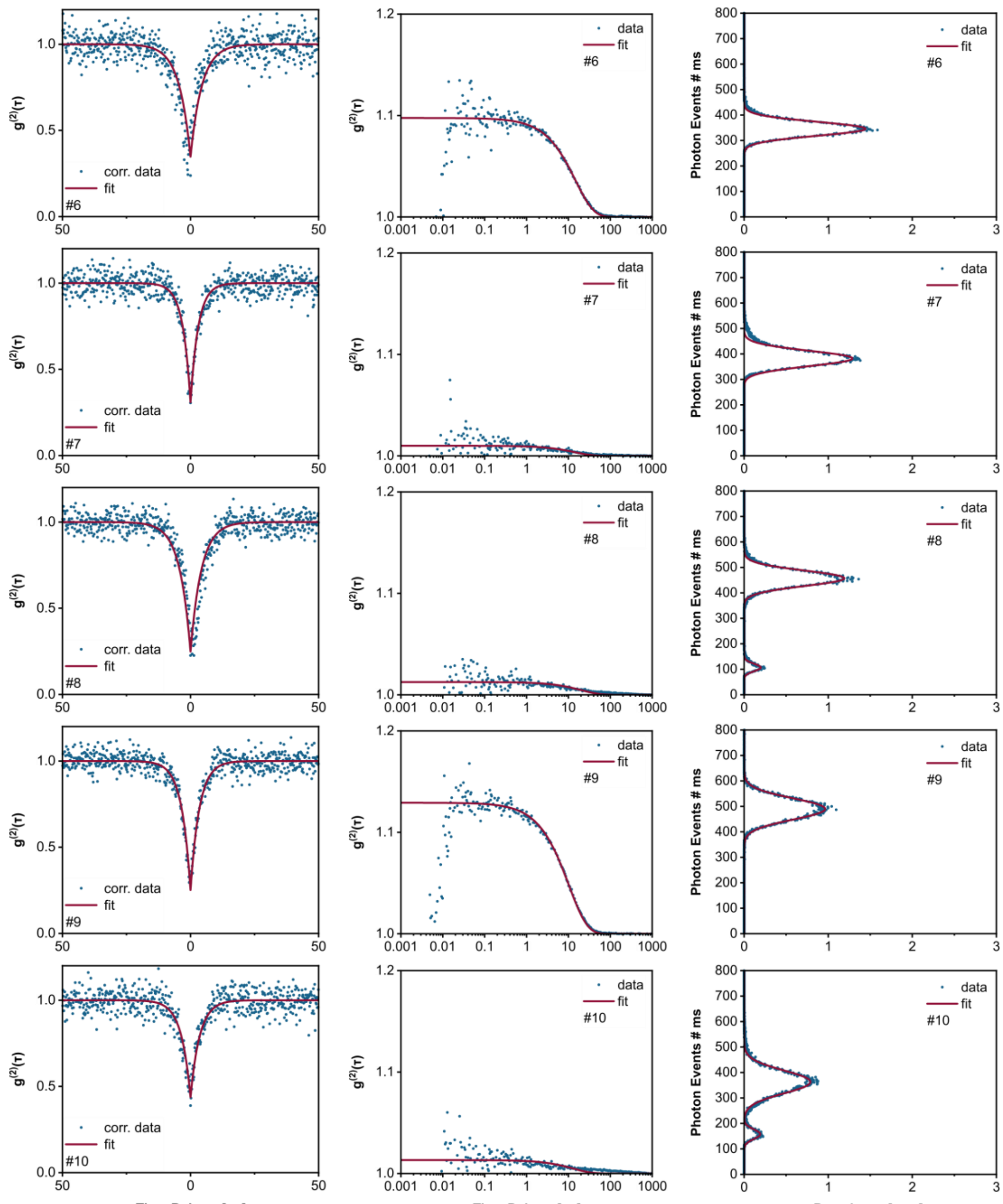

**Figure S4.** Photon correlation and histogram data of individual single TXO-TPA molecules in PMMA. Importantly antibunching data, bunching data and histograms of the photon emission rate were measured during the same lifetime cycle of the molecule. Left column: $g^{(2)}(\tau)$ correlation data in a time delay range between -50 and 50 ns. Shown are background corrected data (Equation SI2) and the corresponding exponential fits (Equation SI3). The antibunching characteristics with $g^{(2)}(0) < 0.5$ clearly confirm the single photon emission character of TXO-TPA. Middle column: $g^{(2)}(\tau)$ correlation data within a time delay of 1000 μs. Shown are the raw data without any post-processing and the corresponding exponential fits (Equation SI4) to the bunching decay. Bunching is present for curves approaching above $g^{(2)}(\tau)$ values larger than one on intermediate time scales. Right column: Histograms of the photon emission rate, derived from intensity time traces recorded during the photon correlation experiments with a time binning of 10 ms. The photon events per time are assigned to the prevalence of measuring the respective photon rate over different time bins. The raw data with the corresponding (multi-)gaussian fits (Equation SI5) are displayed. All measurements were carried out at 532 nm excitation wavelength and 300 μW cw excitation power with circularly polarized light.

## SI 7: Photon Correlation and Histogram Data of TXO-TPA in DPEPO

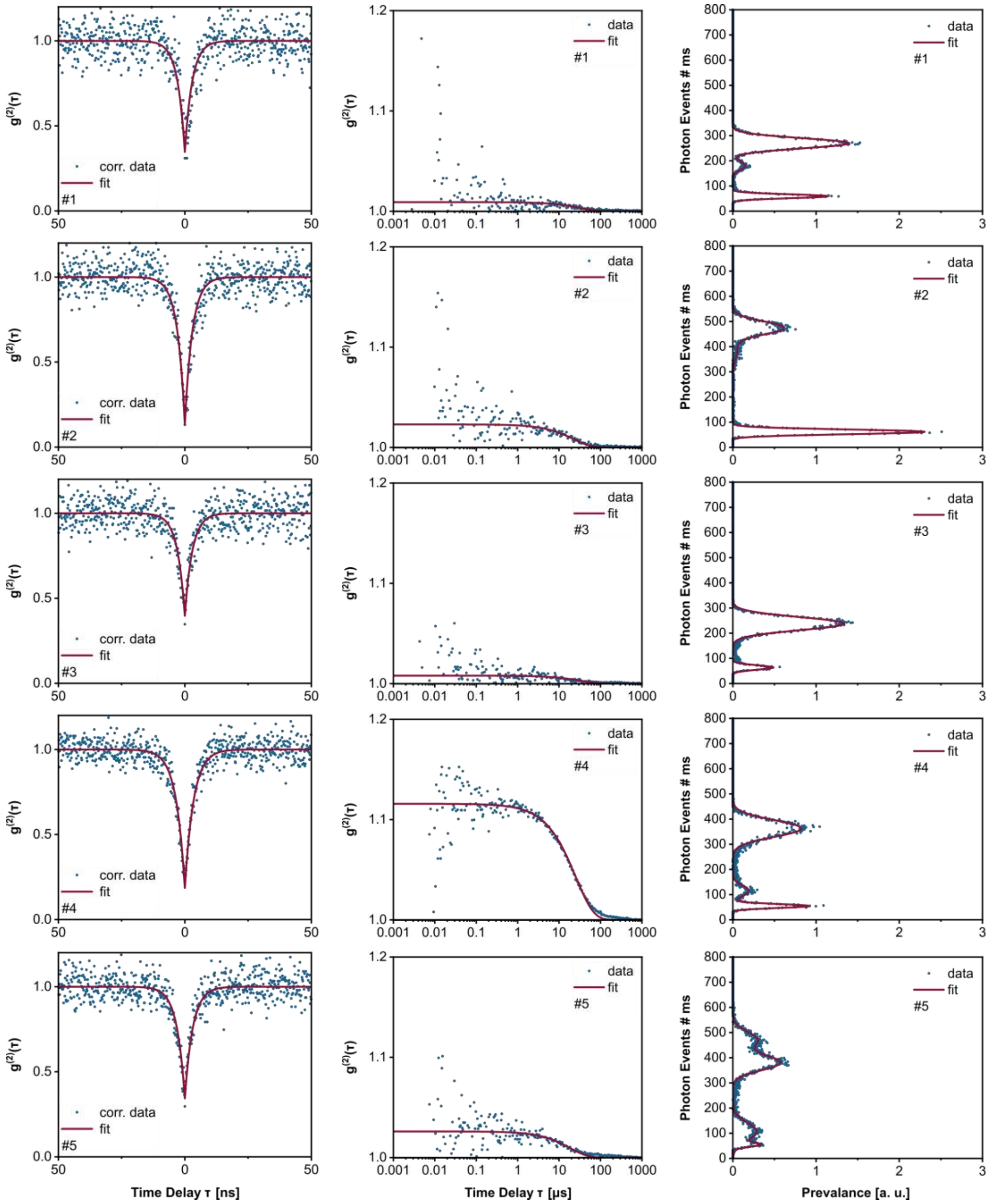

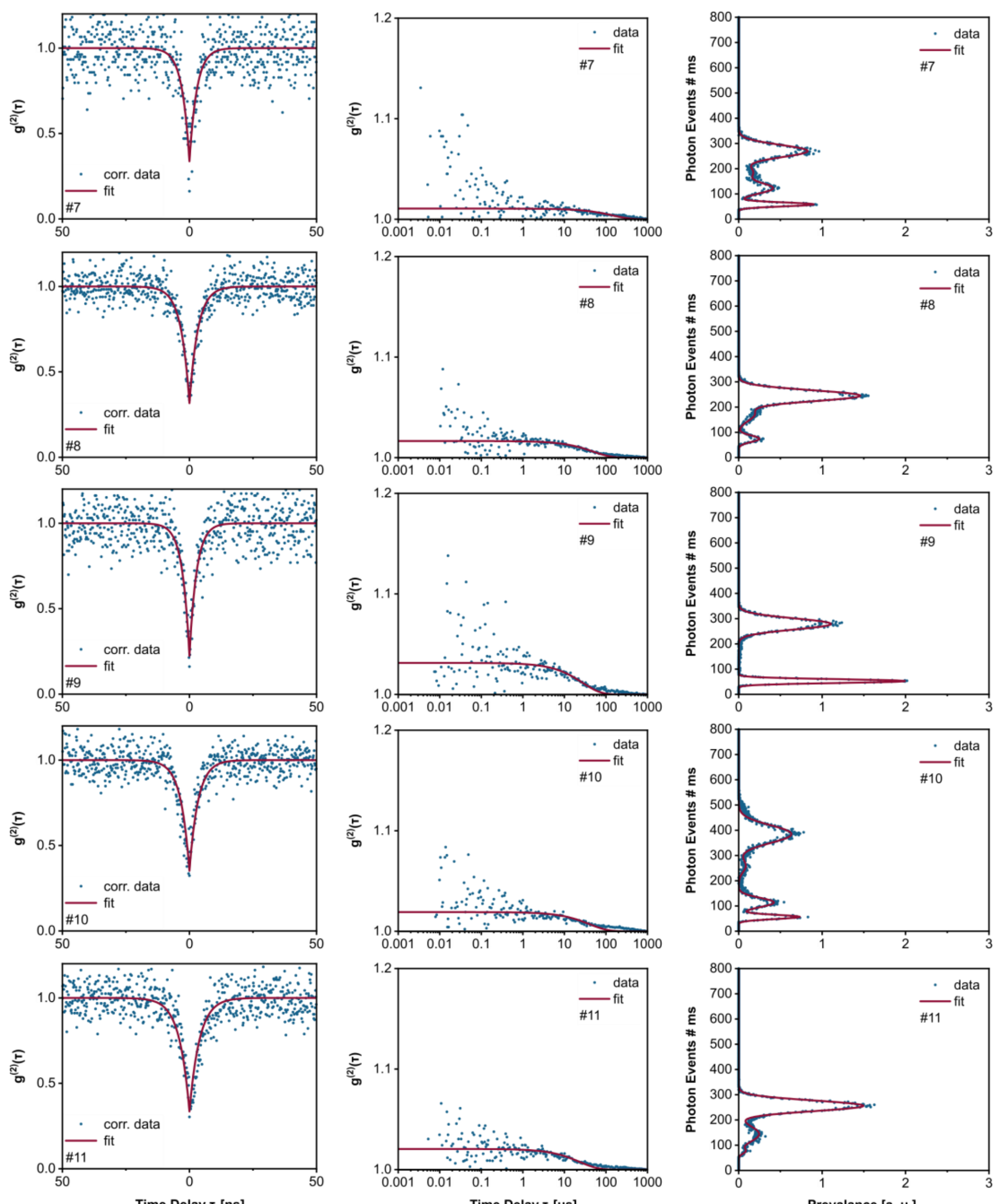

**Figure S5.** Photon correlation and histogram data of individual single TXO-TPA molecules in DPEPO. Importantly antibunching data, bunching data and histograms of the photon emission rate were measured during the same lifetime cycle of the molecule. Left column: $g^{(2)}(\tau)$ correlation data in a time delay range between -50 and 50 ns. Shown are background corrected data (Equation SI2) and the corresponding exponential fits (Equation SI3). The antibunching characteristics with $g^{(2)}(0) < 0.5$ clearly confirm the single photon emission character of TXO-TPA. Middle column: $g^{(2)}(\tau)$ correlation data within a time delay of 1000 µs. Shown are the raw data without any post-processing and the corresponding exponential fits (Equation SI4) to the bunching decay. Bunching is present for curves approaching above $g^{(2)}(\tau)$ values larger than one on intermediate time scales. Right column: Histograms of the photon emission rate, derived from intensity time traces recorded during the photon correlation experiments with a time binning of 10 ms. The photon events per time are assigned to the prevalence of measuring the respective photon rate over different time bins. The raw data with the corresponding (multi-)gaussian fits (Equation SI5) are displayed. All measurements were carried out at 532 nm excitation wavelength and 300 µW cw excitation power with circularly polarized light.

## SI 8: Photon Correlation and Histogram Data of TXO-TPA in UGH-3

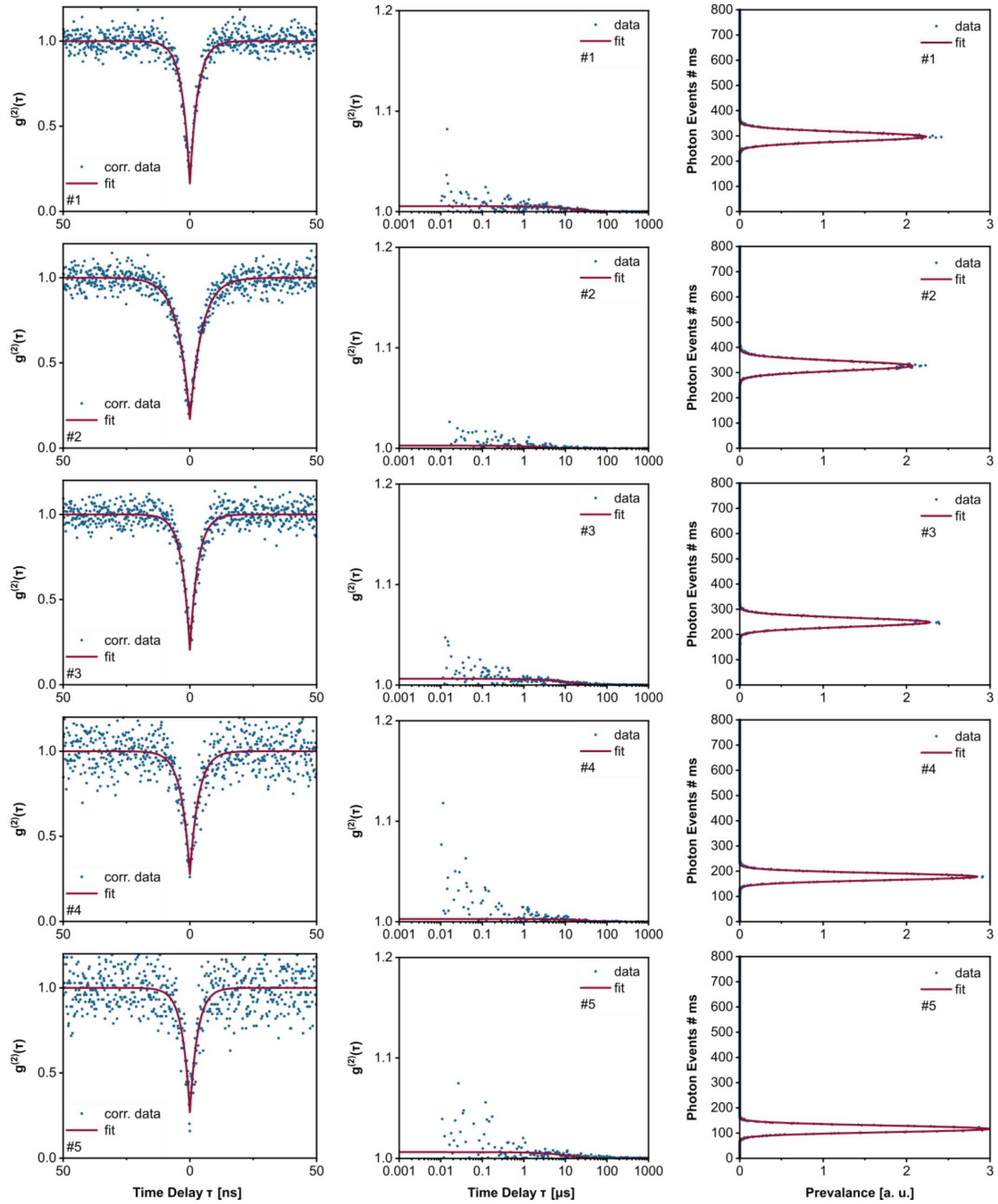

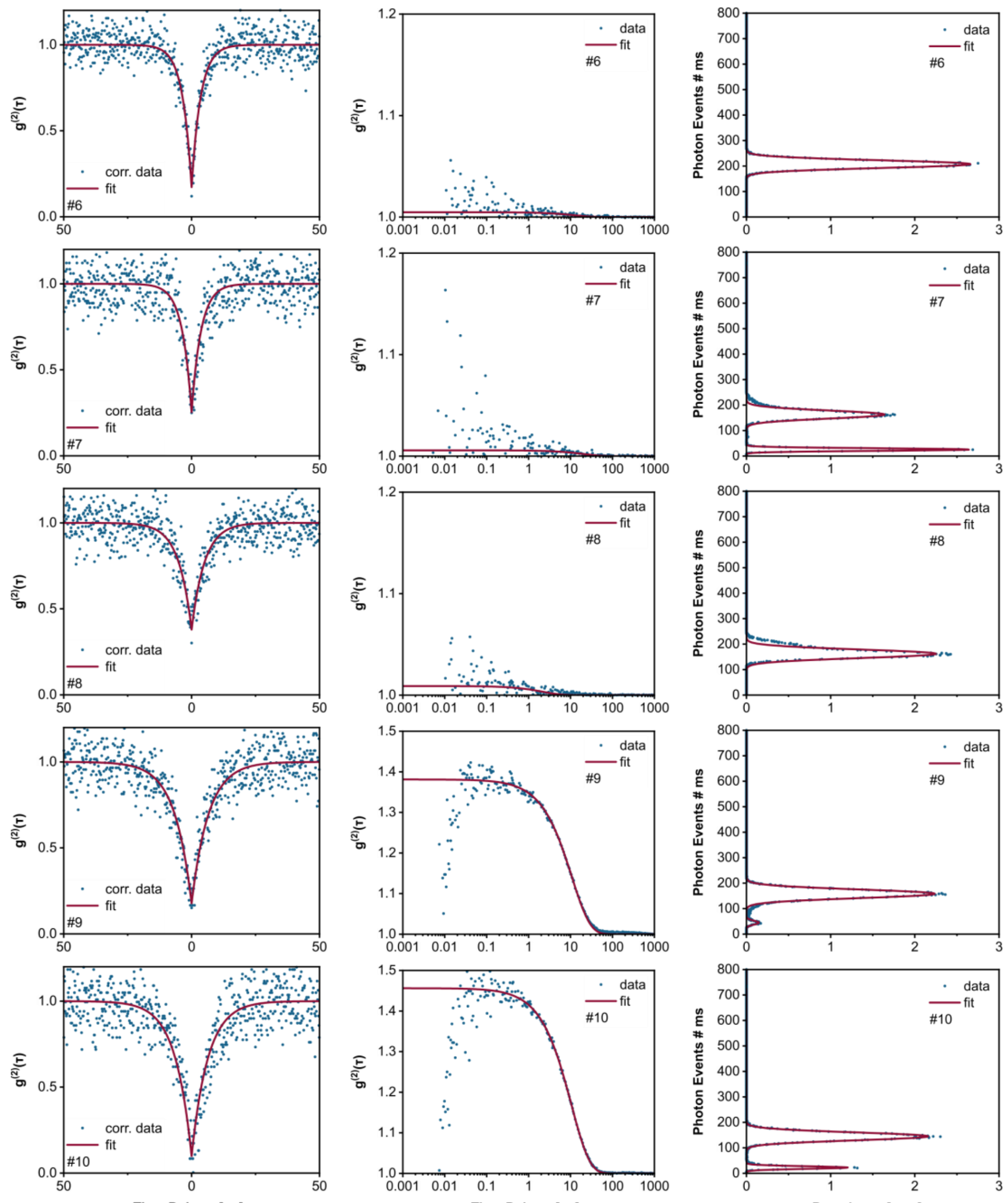

**Figure S6.** Photon correlation and histogram data of individual single TXO-TPA molecules in UGH-3. Importantly antibunching data, bunching data and histograms of the photon emission rate were measured during the same lifetime cycle of the molecule. Left column: $g^{(2)}(\tau)$ correlation data in a time delay range between -50 and 50 ns. Shown are background corrected data (Equation SI2) and the corresponding exponential fits (Equation SI3). The antibunching characteristics with $g^{(2)}(0) < 0.5$ clearly confirm the single photon emission character of TXO-TPA. Middle column: $g^{(2)}(\tau)$ correlation data within a time delay of 1000 µs. Shown are the raw data without any post-processing and the corresponding exponential fits (Equation SI4) to the bunching decay. Bunching is present for curves approaching above $g^{(2)}(\tau)$ values larger than one on intermediate time scales. Right column: Histograms of the photon emission rate, derived from intensity time traces recorded during the photon correlation experiments with a time binning of 10 ms. The photon events per time are assigned to the prevalence of measuring the respective photon rate over different time bins. The raw data with the corresponding (multi-)gaussian fits (Equation SI5) are displayed. All measurements were carried out at 532 nm excitation wavelength and 300 µW cw excitation power with circularly polarized light.

## SI 9: Saturation of the Photon Emission Rate

We have measured the photon emission rate of single TXO-TPA molecules as function of the excitation power to identify a suitable excitation power density for photon correlation experiments and to determine the saturation regime of TXO-TPA within the three different hosts (PMMA, DPEPO and UGH-3). The saturation behavior of the emission rate (intensity $I$) as function of the excitation power $P$ is described by the following equation with the saturation power $P_S$ and the maximum emission rate $I_{\infty}$:

$$I = I_{\infty} \cdot \left(\frac{P/P_S}{1+P/P_S}\right) \qquad \text{(SI6)}$$

The corresponding power seria are shown in Figure S7 with the emission rate of the TXO-TPA molecules corrected by the host contributions, which are depicted in the bottom row.

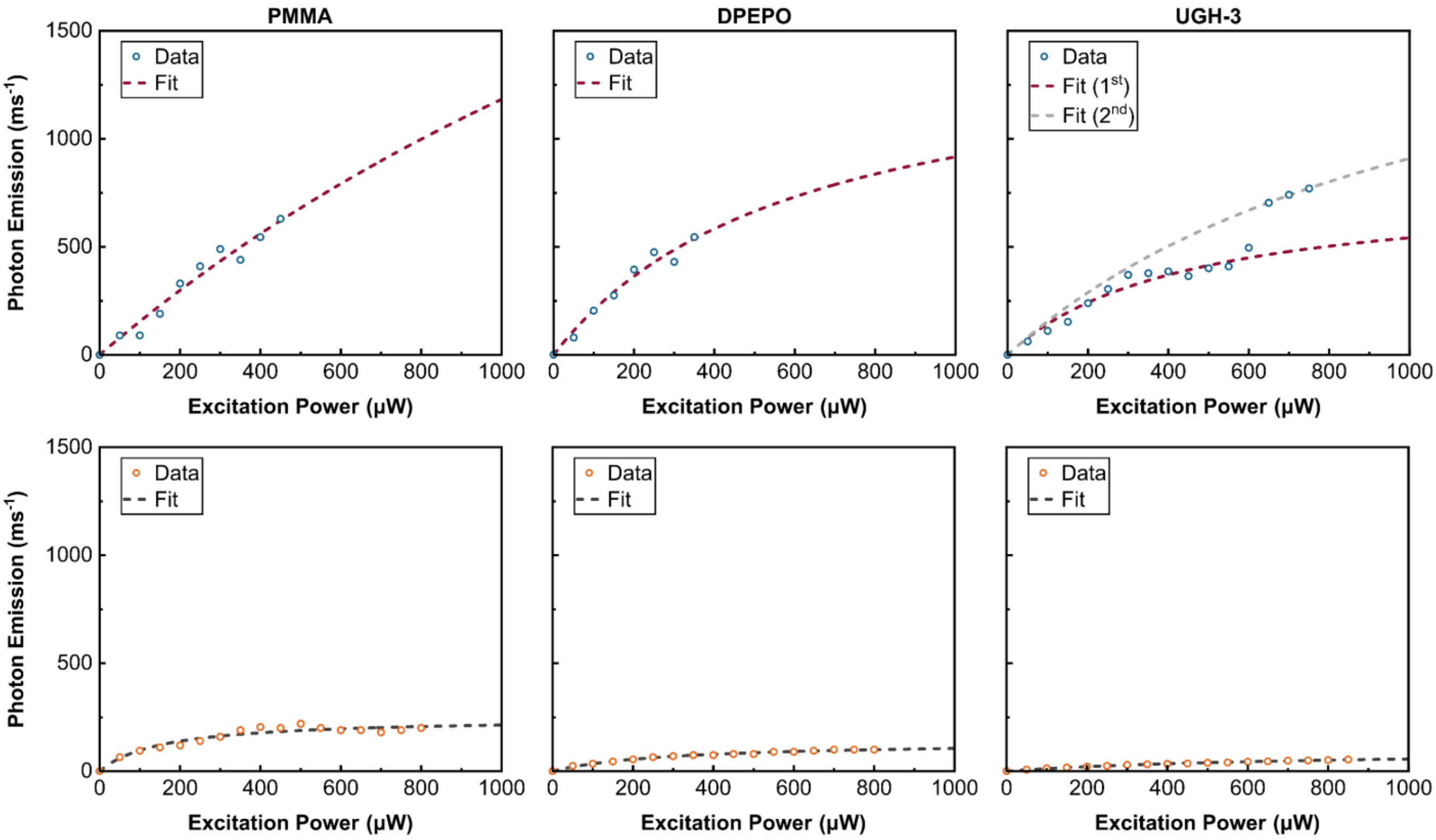

**Figure S7.** Photon emission rate of single TXO-TPA molecules in PMMA, DPEPO and UGH-3 (from left to right) as function of the excitation power. The top row corresponds to the power series of representative TXO-TPA molecules with the photon emission rate corrected by the host contribution. The parasitic emission of the respective host materials was measured beside a single molecule spot (bottom row). Fits based on Equation SI6 are presented by dashed lines. In case of TXO-TPA in PMMA and DPEPO the fits are error-prone as the stability of single molecules in these hosts is limited to about 400 µW excitation power. In case of UGH-3 two different saturation regimes can be identified, which we attribute to dissolvement of the host rigidity at excitation powers above 600 µW. This results in a behavior comparable to the amorphous host materials. All measurements were carried out at 532 nm excitation wavelength with circularly polarized light. The excitation power was adjusted with a continuous grey filter wheel and measured with a power meter.

The parasitic host contribution is highest in PMMA, and lowest in UGH-3. In case of PMMA and DPEPO the fits are error-prone as the stability of single TXO-TPA molecules in these host materials is limited to an excitation power of about 400 μW. The fits should thus be taken more as illustrative guide. In case of UGH-3 two different saturation regimes can be identified, which we attribute to dissolvement of the host rigidity at excitation powers above 600 μW. Interestingly the second saturation regime is then comparable to the non-rigid host materials. The uniform excitation power of 300 μW throughout the photon correlation measurements was chosen for several reasons: 1.) An adequate signal-to-background ratio is to be ensured, which is the case at moderate and high excitation powers as the parasitic host contribution is saturating. 2.) Excitation of TXO-TPA close to saturation. 3.) Stability of the emitter entity during correlation experiments.

## SI 10: Transient PL of TXO-TPA in the Emitter Ensemble

Ensemble thin film samples of TXO-TPA dispersed in PMMA, DPEPO and UGH-3 at an emitter concentration of 1 wt% were prepared (see experimental section). The confocal setup described in the experimental section was used for time-correlated single photon counting (TCSPC). A pulsed laser ($\lambda_{ex}$ = 520 nm) and a hardware correlator (PicoHarp 300) synchronized by the electrical signal of the laser, together with an avalanche photodetector (Excelitas SPCM-AQRH-14, QE 65 % at 650 nm, dark counts < 100 counts·$s^{-1}$) were utilized to record histograms. The prompt and delayed decay were measured with a laser pulse width of ~ 4 ns and ~ 36 ns, respectively. The instrument response function was measured with a cover glass at the sample position.

Lifetime density analysis was applied to analyze the prompt decay of TXO-TPA. A pseudo-continuous distribution of decay times is assumed:

$$I(t) = \sum_{\tau_i} a(\tau_i) \cdot e^{-\frac{t}{\tau_i}} * IRF(t) \qquad \text{(SI7)}$$

Here $I(t)$ is the photoluminescence signal, $a(\tau_i)$ are the amplitudes of the respective decay constants $\tau_i$ and $IRF(t)$ is the instrument response of the setup.

The prompt fluorescence decay together with the lifetime density distributions are shown in Figure S8. The mean fluorescence lifetimes are deduced from the maxima in the lifetime density distribution. The occurrence of three distinct maxima for the rigid UGH-3 host refers to a pronounced distribution of prompt lifetimes (0.5 ns, 2.8 ns and 21.1 ns) over the ensemble. These results are complementary to the antibunching measurements which revealed a distribution of the antibunching width in case of UGH-3. In PMMA and DPEPO the dominant lifetimes are 21.7 ns and 22.4 ns, respectively and are comparable to the long component in UGH-3. A less pronounced fast component of 1.4 ns can be identified in PMMA and DPEPO.

The transient PL measurements of the delayed component are displayed in Figure S9. An exponential decay was fitted to retrieve the mean lifetime constant of the delayed fluorescence. The delayed lifetime is 13 µs in PMMA, 10 µs in DPEPO and 8 µs in UGH-3, which reveals a strong impact of the host material on the rISC/ISC dynamics. The ensemble data lack the direct information on local host-emitter interactions, which we gained from single molecule spectroscopy.

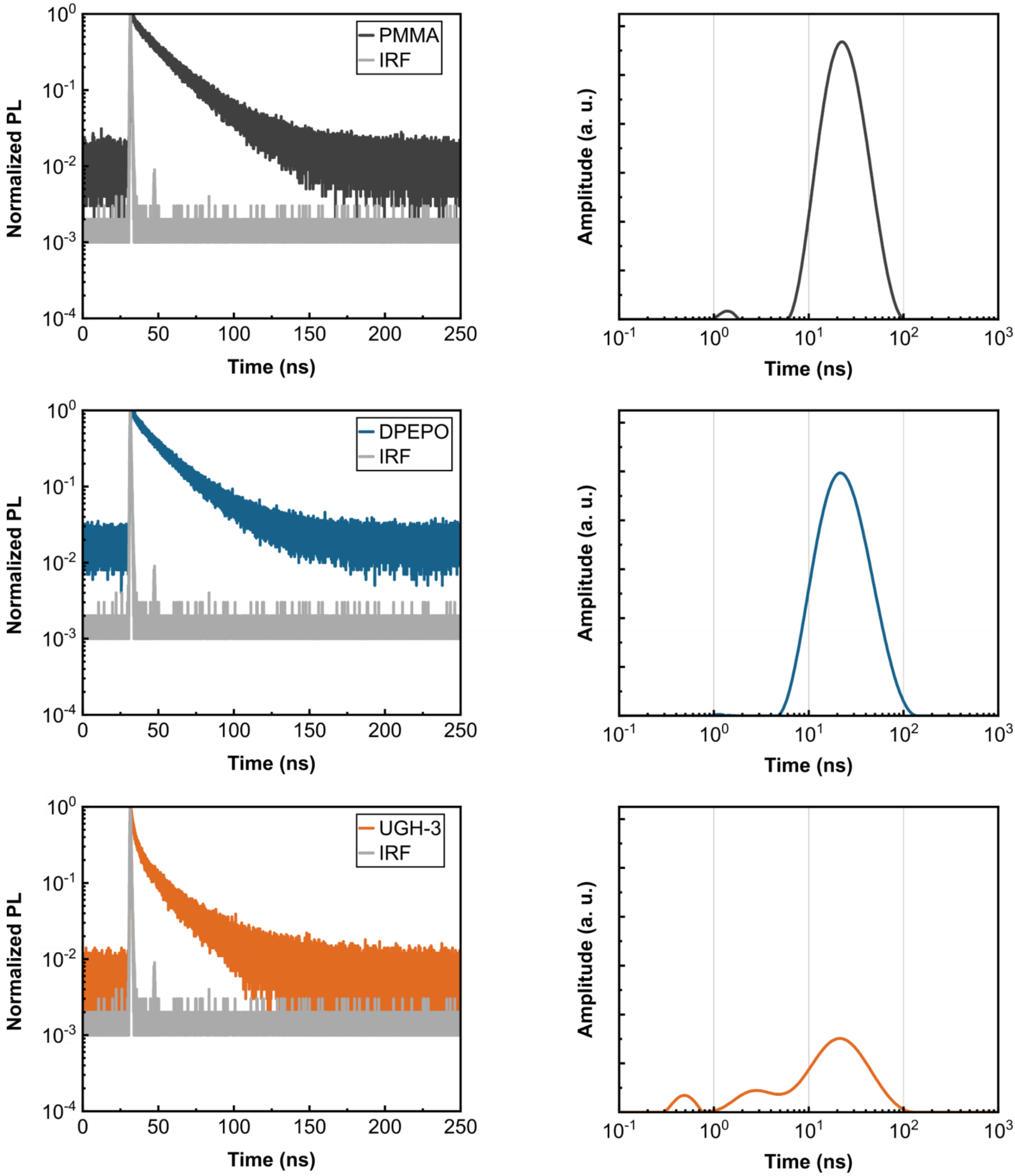

**Figure S8.** Transient PL measurements of the prompt fluorescence of TXO-TPA dispersed at 1 wt% in PMMA, DPEPO and UGH-3 (from top to bottom). Left column: Normalized intensity histograms with the instrument response function (IRF) shown in grey. Right column: Corresponding lifetime density distributions (Equation SI7) of the prompt decay component.

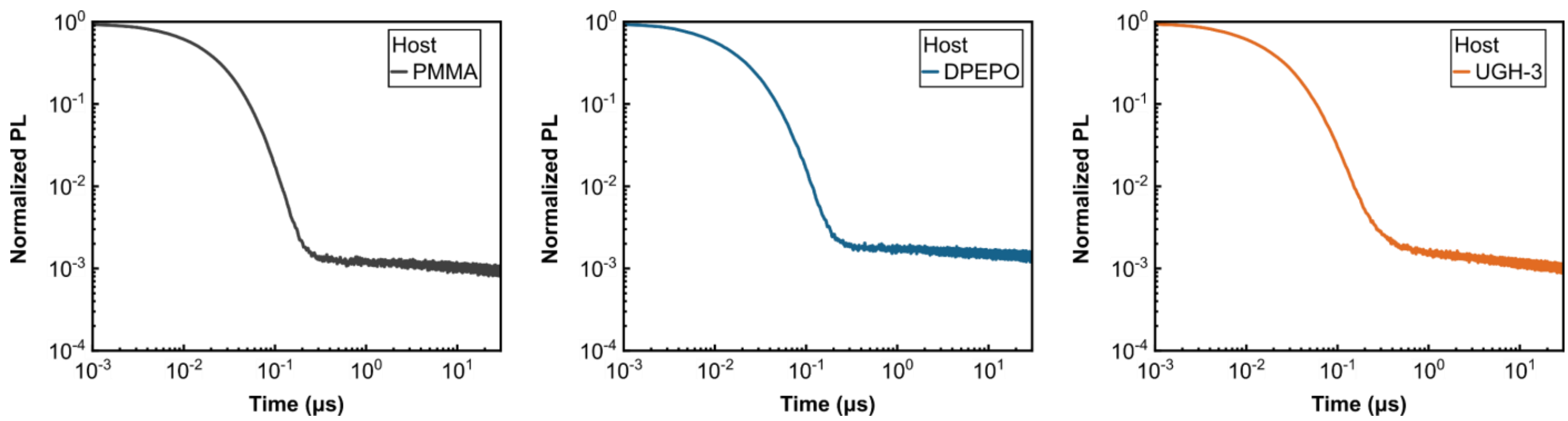

**Figure S9.** Transient PL measurements of the delayed fluorescence of TXO-TPA dispersed at 1 wt% in PMMA, DPEPO and UGH-3 (from left to right). Shown are the normalized intensity histograms after applying a moving average filter to the raw data.